\documentclass[11pt]{article}
	
	\newcommand{\blind}{0}
	
    \makeatletter
    \renewcommand\section{\@startsection {section}{1}{\z@}%
                                       {-3.5ex \@plus -1ex \@minus -.2ex}%
                                       {2.3ex \@plus.2ex}%
                                       {\normalfont\fontfamily{phv}\fontsize{16}{19}\bfseries}}
    \renewcommand\subsection{\@startsection{subsection}{2}{\z@}%
                                         {-3.25ex\@plus -1ex \@minus -.2ex}%
                                         {1.5ex \@plus .2ex}%
                                         {\normalfont\fontfamily{phv}\fontsize{14}{17}\bfseries}}
    \renewcommand\subsubsection{\@startsection{subsubsection}{3}{\z@}%
                                        {-3.25ex\@plus -1ex \@minus -.2ex}%
                                         {1.5ex \@plus .2ex}%
                                         {\normalfont\normalsize\fontfamily{phv}\fontsize{14}{17}\selectfont}}
    \makeatother
	\usepackage{amsmath}
	\usepackage{graphicx}
	\usepackage{enumerate}
	\usepackage{xcolor}
	\usepackage{natbib} %comment out if you do not have the package
	\usepackage{url} % not crucial - just used below for the URL
        \usepackage[utf8]{inputenc}
    \usepackage{setspace}
    \usepackage[margin=1.0in]{geometry}
    \usepackage{amssymb}
    \usepackage{amsthm}
    \usepackage{colordvi,verbatim,hyperref}
    \usepackage{color}
    \usepackage{caption}
    \usepackage{algorithm,algorithmic}
    \usepackage{subcaption}
    \usepackage[justification=centering]{caption}
    \usepackage{authblk}
    \usepackage[toc,page]{appendix}
    \usepackage{lineno}
    \usepackage{tablefootnote}
        \allowdisplaybreaks

\begin{document}
		
			%%%%%%%%%%%%%%%%%%%%%%%%%%%%%%%%%%%%%%%%%%%%%%%%%%%%%%%%%%%%%%%%%%%%%%%%%%%%%%
		\def\spacingset#1{\renewcommand{\baselinestretch}%
			{#1}\small\normalsize} \spacingset{1}
		%%%%%%%%%%%%%%%%%%%%%%%%%%%%%%%%%%%%%%%%%%%%%%%%%%%%%%%%%%%%%%%%%%%%%%%%%%%%%%
		
		\if0\blind
		{
\title{\textbf{A Transdisciplinary Approach for Generating Synthetic but Realistic Domestic Sex Trafficking Networks}}
\author[1]{Daniel Kosmas\footnote{Corresponding author (d.kosmas@northeastern.edu).}}
\author[2]{Christina Melander}
\author[3,4]{Emily Singerhouse}
\author[5]{Thomas C. Sharkey} 
\author[1]{Kayse Lee Maass}
\author[2]{Kelle Barrick}
\author[6]{Lauren Martin}
\affil[1]{\footnotesize Department of Mechanical and Industrial Engineering, Northeastern University, Boston, MA 02115, USA}
\affil[2]{\footnotesize Division of Applied Justice Research, RTI International, Research Triangle Park, NC 27709, USA}
\affil[3]{\footnotesize Strategic Prevention Solutions, Juneau, AK 99802, USA}
\affil[4]{\footnotesize Singerhouse Research Consulting LLC, Tampa, FL 33510, USA}
\affil[5]{\footnotesize Department of Industrial Engineering, Clemson University, Clemson, SC 29634, USA}
\affil[6]{\footnotesize School of Nursing, University of Minnesota,  Minneapolis, MN 55455, USA}
\date{}
\maketitle
   
		} \fi
		
		\if1\blind
		{

            \title{\textbf{A Transdisciplinary Approach for Generating Synthetic but Realistic Domestic Sex Trafficking Networks}}
			\author{}
			\date{}
\maketitle

		} \fi
		\bigskip

\vspace{-1.5cm}
\begin{abstract}
    One of the major challenges associated with applying operations research (OR) models to disrupting human trafficking networks is the limited amount of reliable data sources readily available for public use, since operations are intentionally hidden to prevent detection, and data from known operations are often incomplete. To help address this data gap, we propose a network generator for domestic sex trafficking networks by integrating OR concepts and qualitative research. Multiple sources regarding sex trafficking in the upper Midwest of the United States have been triangulated to ensure that networks produced by the generator are realistic, including law enforcement case file analysis, interviews with domain experts, and a survivor-centered advisory group with first-hand knowledge of sex trafficking. The output models the relationships between traffickers, so-called ``bottoms", and victims. This generator allows operations researchers to access realistic sex trafficking network structures in a responsible manner that does not disclose identifiable details of the people involved. We demonstrate the use of output networks in exploring policy recommendations from max flow network interdiction with restructuring. To do so, we propose a novel conceptualization of flow as the ability of a trafficker to control their victims.  Our results show the importance of understanding how sex traffickers react to disruptions, especially in terms of recruiting new victims.
\end{abstract}
	\noindent%

	%\newpage
	\spacingset{1.5} % DON'T change the spacing!

\section{Introduction}
\label{sec:intro}
Human trafficking is human rights abuse documented both in the United States and abroad. Although the magnitude of the problem is difficult to determine \citep{farrell2020measuring, fedina2015use}, estimates from the International Labor Organization suggest that nearly 24.9 million people were victims of human trafficking in 2016 \citep{international2017global}. Human trafficking involves the use of force, fraud, or coercion in order to exploit a person for the purposes of labor or services (labor trafficking) or sexual exploitation (sex trafficking) \citep{carpenter2016nature, martin2014benefit, polaris2017typology, preble2019under}. Despite the growing awareness of the problem, there has been limited research into developing quantitative tools to address the problem of human trafficking \citep{caulkins2019call, dimas2021survey, konrad2017overcoming}. This is partially due to a lack of readily-usable data to appropriately populate these tools. Since these networks are illicit in nature, their operations are hidden to avoid detection \citep{farrell2020measuring, fedina2015use, konrad2017human, konrad2017overcoming}. Current data for analytic approaches often comes from case files \citep{xieinterdependent}, scraped web data on sex advertisements \citep{keskin2021cracking}, or data pertaining to massage parlors \citep{mayorga2019countering, white2021you}. This data can be viewed as the ``forward-facing" aspects of the operations of sex trafficking networks in terms of how they interact with the outside world, thus it is (somewhat) publicly accessible. However, there are data gaps around the more internal-facing operations and social connections between traffickers and victims. We seek to develop a way to generate data that can help fill some of these gaps to enable quantitative research on human trafficking networks. In order to highlight the importance of our data collection methods, we first survey the background on sex trafficking networks (Section \ref{sec:1.1}) and OR modeling of them (Section \ref{sec:1.2}).  We then position our data within the broader landscape of data on sex trafficking (Section \ref{sec:1.3}) and highlight our contributions (Section \ref{sec:1.4}).

\subsection{\emph{Background on Sex Trafficking Networks}} \label{sec:1.1}
Analytical tools that seek to understand and disrupt the operations of sex trafficking networks could be quite powerful, assuming their consequences are analyzed by domain experts. We can consider sex trafficking networks as mathematical networks. Networks consist of nodes, representing the entities, and arcs, representing the connections between nodes. A common example is a social network, where the nodes represent people and the arcs represent the different connections between people. Many models in operations research (OR) involve networks, since they can be used to model the connections between entities and how people and goods can move between different locations or states. Operations researchers have begun to explore how network models can be used to disrupt sex trafficking networks, and have experienced difficulty in obtaining high quality data \citep{dimas2021survey, mayorga2019countering, tezcan2020human}.  The focus on describing sex trafficking networks tends to be on empirical studies that lack the type of data needed for operations research modeling, as we discuss in the remainder of this subsection.  

\cite{cockbain2018offender} provides a comprehensive analysis of networks of victims and trafficking in case file records from six law enforcement investigations into domestic minor sex trafficking the United Kingdom (UK). She analyzes police operational files, court records, prosecution case files, and interviews with convicted traffickers, police investigators and prosecutors to explore the demographics of the traffickers and victims in these cases and the tactics used by traffickers to recruit and retain victims. She produces the social networks of the victims and of the traffickers from these analyses. However, her work does not discuss how to generalize the structures of these networks to systematically create different, yet plausible and realistic, instances of a domestic sex trafficking network. In particular, there is no information on the connections between victims and traffickers.

\cite{dank2014estimating} conducted a comprehensive analysis of the size and structure of the commercial sex economy in eight major cities in the United States of America (USA). As part of this study, they investigated the operations of sex trafficking, as well as how victims are recruited and managed. They also qualitatively describe how traffickers are socially connected to share information. They noted, from interviews with both traffickers and law enforcement, that traffickers are highly networked socially, but rarely form business partnerships. This work helps to set the foundation for creating realistic domestic sex trafficking networks.

\cite{veldhuizen2021madams} explore sex trafficking networks with female traffickers by analysing publicly available federally prosecuted case files. Their work identifies operational characteristics, including the number of victims and traffickers, their ages, and whether or not victims are domestic or international, for 44 sex trafficking operations in the US. They also explore the roles and duties that women performed in sex trafficking networks. Their work does not, however, include the social and operational connections between traffickers and victims.

\cite{mancuso2014not} investigated how a person's role in a sex trafficking network impacted their centrality measures by analyzing one trafficking network obtained from one case file that spanned Italy to Nigeria. \cite{campana2016structure} further explored the structure of that sex trafficking network, focusing on when multiple members of the network participated in the same trafficking events. We augment the social network perspective by incorporating the insights of domain experts and people with lived experiences in sex trafficking networks, elucidating hidden aspects of domestic sex trafficking networks that cannot be determined by advertisements or law enforcement investigations.

\subsection{\emph{Background on OR Modeling of Sex Trafficking Networks}} \label{sec:1.2}
OR models for societal challenges can be supplemented with insights from domain experts from a wide array of disciplines, such as qualitative researchers. Transdisciplinary research is necessary to conduct appropriate research related to human trafficking networks and seeks to address complex societal challenges through the integration of knowledge and methods of different disciplines \citep{lotrecchiano2018transdisciplinary}. Because of the complexities of the lived experiences of survivors of sex trafficking, domain expertise and effective communication \citep{martin2022comms} is necessary to ensure that any analytical tools developed for the purpose of understanding and disrupting sex trafficking networks appropriately consider the human element of these networks. There are many challenges with building a transdisciplinary research team, but the time and effort put in to build such a team can result in scholarly works better grounded in the application area \citep{sharkey2021better}. We have applied a transdisciplinary approach in creating the proposed network generator by collaborating with domain experts, who have been investigating human trafficking for over $10$ years (see \cite{barrick2015developing, barrick2022estimating, barrick2014farmworkers, barrick2021law, martin2014benefit, martin2017mapping, martin2014mapping}), and a survivor-centered advisory group.

Incorporating domain expertise in the application of OR is critical since it allows the created models to focus on the true underlying problems faced in the application area.  This is especially important when the system cannot be directly observed.  \cite{morris1967art} discusses the types of tasks necessary to inform models while \cite{willemain1994insights, willemain1995model} discuss the process by which experts create models when faced with a practical problem.  However, less research has been done on how to integrate domain expertise, both within other academic disciplines and from practitioners, on socially-sensitive, hidden issues, like human trafficking. \cite{caulkins2019call} discuss the potential types of insights that can be obtained by using engineering models to understand human trafficking and discuss the need to partner with human trafficking domain experts. \cite{sharkey2021better} present a high-level approach for how to integrate this expertise into the modeling process, where they observe that data is one of the key areas where experts need to inform the modeling process.%.  A critical observation of \cite{sharkey2021better} is that data is one of the key areas where experts need to inform the modeling process.  

\cite{dimas2021survey} provides a comprehensive review of existing literature in applying operations research and analytics to the understanding and disruption of human trafficking. Researchers are currently considering investigating two different perspectives for sex trafficking networks. One perspective considers how sex trafficking networks interact with different locations. \cite{mayorga2019countering} used web scraping from an online escort service to populate a network interdiction model, where the victims in the trafficking network are considered the flow being moved from location to location. \cite{keskin2021cracking} proposed a process of grouping online advertisements to predict the movement of sex trafficking networks based on where related advertisements appear. \cite{white2021you} used web scraping to gather data on illicit massage parlors, and determines potential contributing factors for where illicit massage parlors may be located. These works use a visible, forward-facing part of the sex trafficking networks, the advertisements, to perform their analysis. Yet, advertisements are not the same as number of victims or people in the network and it is unclear exactly how advertisements map to the underlying phenomena of trafficking itself. The other perspective considers the relationships of the individuals within the sex trafficking network. \cite{cockbain2018offender} performs social network analysis of traffickers and victims in six sex trafficking operations in the UK that law enforcement disrupted. 

Network interdiction has received considerable attention for its ability to assist in the disruption of illicit trafficking, including nuclear smuggling \citep{morton2007models} and drug trafficking \citep{malaviya2012multi}. As such, it has been suggested to be a tool that may prove useful in combating human trafficking \citep{smith2020survey}. There has been limited work in applying network interdiction to human trafficking networks. \cite{mayorga2019countering} discusses a robust max flow network interdiction model, where there is uncertainty about the capacities of arcs. \cite{tezcan2020human} consider a multi-period max flow network interdiction model where there is uncertainty about the success of the proposed interdictions, and the success of interdictions in each time period is dependent on the success of interdictions in the previous time periods. The model is unique in that they consider flow to be the desirability of a trafficker to travel along different routes. These models both model the movement of people across geographic areas. \cite{xieinterdependent} propose a network interdiction model on interdependent networks, building off of the work of \cite{baycik2018interdicting}. Their model considers flow to be the victims in the network, being transported between people with different roles in the sex trafficking network. These models all focus on larger scale operations, and all fail to consider the autonomy of the victims themselves in the trafficking networks. The conceptualization we propose treats victims as more than a product to be moved through the network and, importantly, directly captures the control that exists within trafficking networks.

\subsection{\emph{Framing Data Sources on Sex Trafficking Networks}} \label{sec:1.3}

\begin{figure}[h]
    \centering
    \includegraphics[width=0.65\linewidth]{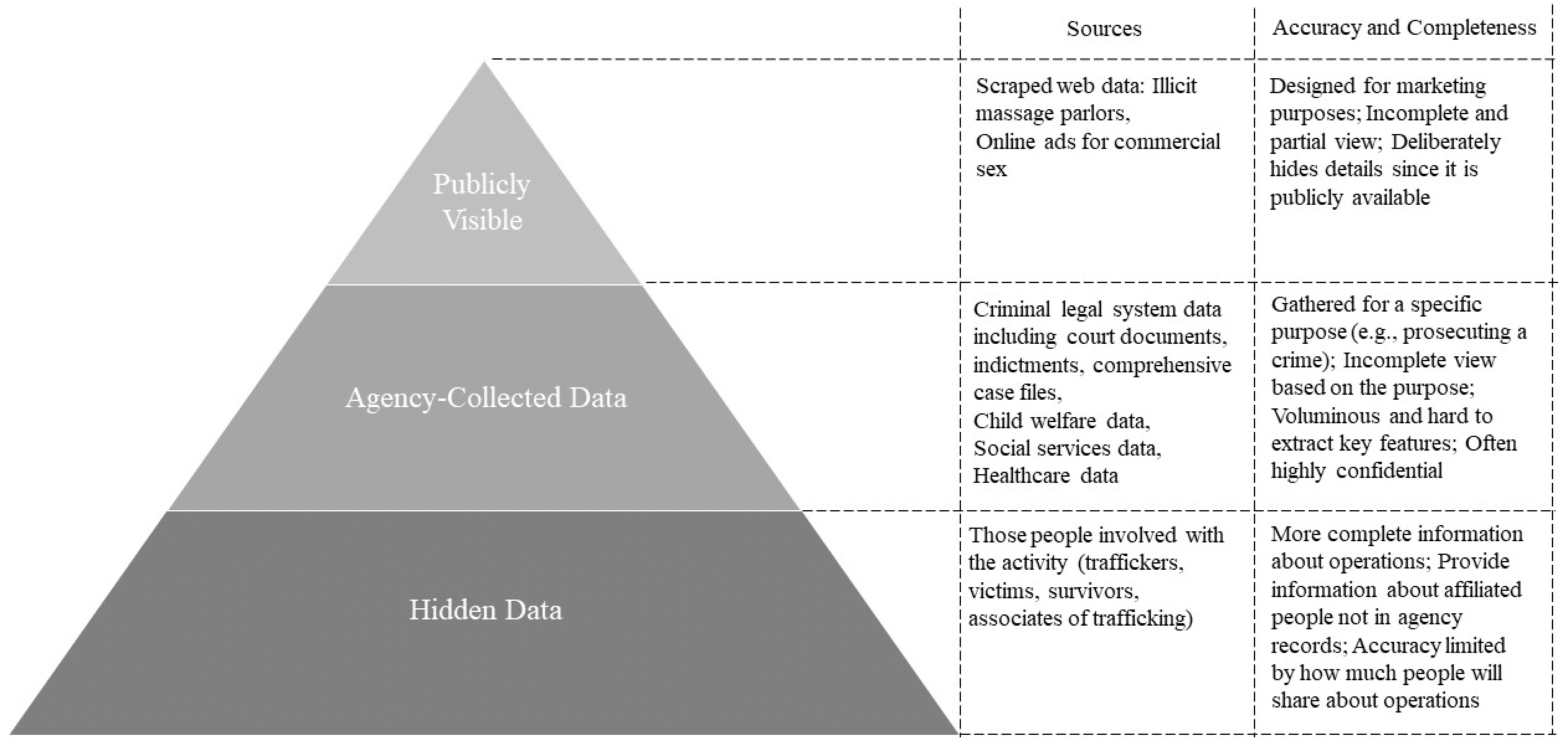}
    \caption{Types of human trafficking data for use in OR models}
    \label{fig:othersources}
\end{figure}

Data sources that can provide sex trafficking network details include (but are not limited to) publicly available data (e.g., web advertisements), secondary agency-collected data (e.g., law enforcement and service providers case files) and primary (hidden) data (e.g., interviews with those who have been a part of, or victimized by, these networks). However, each source has limitations, and no single source will have the full information necessary to perform a proper analysis of the network structure; we highlight the differences in these sources of data in Figure \ref{fig:othersources}. Although some data on sex trafficking is publicly available, it is limited to aggregate statistics, such as the number and characteristics of victims identified and perpetrators arrested, and does not provide detailed information about other actors in trafficking networks \citep{motivans2018}, or requires significant effort to extract the data for use in analysis.

In general, agency-collected data is used by agencies and organizations to achieve their internal needs. For example, in the case of  law enforcement, the focus of information included in investigative and prosecutorial case files is on proving the elements of a specific crime. Details that do not help reach this goal but would be relevant to construct the full extent of the trafficking network may be viewed as irrelevant and thus not documented (e.g., information about associates who support trafficking operations without engaging in criminal behavior, such as providing transportation, housing, or childcare). Law enforcement data is further limited in that, by definition, it only includes sex trafficking cases that are known to and officially recorded by law enforcement \citep{cockbain2020using}. Those cases may systematically differ from those that do not come to the attention of law enforcement such that victims with certain demographic characteristics may be more likely to self-identify as a victim and law enforcement may be more proactive in identifying victimization in certain sectors (e.g., targeted stings at massage parlors). Because sex trafficking is known to be under-identified \citep{farrell2019}, the difference between reported and unreported cases could be even more substantial than other crime types. Additionally, extracting the relevant information from the case files to produce the network takes a significant amount of time. Agencies organize and store information in a manner that is most useful for their purposes, not for research. Case files may be incredibly large and include various types of documents, such as handwritten notes, evidence logs, transcripts from victim, suspect, or witness interviews, among others. Sorting, reviewing, and extracting the relevant pieces of information is cumbersome and time consuming.

Primary data, or hidden data, collected from those involved in trafficking operations can help fill some of the gaps in secondary data. Traffickers and survivors of sex trafficking, as well as individuals who work with these populations, are valuable sources of information on sex trafficking networks, but there are numerous challenges associated with gathering data from them. For example, there are logistical difficulties in simply gaining access to traffickers and survivors and ethical issues around asking survivors to recount their exploitation, which can be retraumatizing. Although interviews provide an opportunity to gather complex and nuanced information about trafficking operations, they tend to rely on small, local convenience samples which may limit the generalizability of the data \citep{cockbain2020using, gerassi2017design, weitzer2014new}.

Prior research on trafficking networks has yielded foundational information on the social connections between victims and traffickers through the review of case files and/or interviews with individuals involved in trafficking operations \citep{cockbain2018offender, dank2014estimating, mancuso2014not, campana2016structure, xieinterdependent} and on the locations and movement of sex trafficking operations through the web scraping of commercial sex advertisements \citep{mayorga2019countering, keskin2021cracking, white2021you}. This extant research includes both micro-level explorations of specific trafficking networks, with limitations on generalizing the findings to other networks, and higher-level descriptive information on operational characteristics (e.g., number of victims and traffickers and roles in the network), without producing network structures. We build on and bridge this prior work by incorporating the insights of domain experts and people with lived experience to better understand the full variation in how domestic sex trafficking networks realistically operate.% in the real world.

\subsection{\emph{Our Contributions}} \label{sec:1.4}

We propose a novel network generator that outputs network configurations that are representative of domestic U.S.\ sex trafficking networks in the real world and that account for realistic variation among operations. This generator takes the number of desired trafficking operations for use in analysis as an input, then produces the operational and social connections between the participants in the network.  Our generator is the product of triangulating multiple data sources with the context of the upper Midwest US, including case file analysis, interviews with people who have domain expertise, and validation with a survivor-centered advisory group. The networks produced by the generator can be used for various analytical models without the need for researchers to collect and validate their own data. Acquiring and cleaning data regarding sex trafficking networks is costly, so this generator will help reduce the barrier to entry for OR researchers to develop and test models to analyze and disrupt sex trafficking. We note that the development process for the network generator is iterative. This current iteration was developed in a specific context (domestic trafficking in the upper-Midwest USA) and will require additional collaboration to expand. The more the generator is used alongside domain experts, the more it can be refined, expanded and further validated for improved accuracy in more and varied contexts.

Additionally, we propose a novel conceptualization of flow for sex trafficking networks; we model flow as the ability of the traffickers to control their victims. Thus, the maximum flow through the network is the total number of victims the traffickers victimize at a particular moment in time. We demonstrate how max flow network interdiction can be applied to these networks and discuss policy recommendations for when the traffickers are able to restructure their operations after the interdiction decisions have been implemented.

The paper is organized as follows. Section \ref{sec:develop} outlines our objectives when developing the network generator and the process of triangulating data sources to develop the generator. Section \ref{sec:output} addresses the assumptions of the network generator and analyzes sample outputs. Section \ref{sec:caseStudy} focuses on a responsible use case study of the generator that applies a novel network interdiction model to disrupt the operations of generated sex trafficking networks. Section \ref{sec:limit} discusses limitations of our network generator and proposes directions for future qualitative research to improve the applicability of the network generator. Section \ref{sec:conc} concludes the paper and discusses directions for future research.

\section{Development of the Network Generator}
\label{sec:develop}

The generator focuses on the social and operational connections of the people within domestic sex trafficking networks with specific operational characteristics. We recognize that these networks are abstractions of the true operations of sex trafficking; the proposed network generator cannot appropriately capture all of the complexities of the lived experiences of trafficking victims and survivors, nor can it account for the human rights abuses and violence that occur in sex trafficking networks. This is a limitation of any analytical approach to understanding the issue of sex trafficking.

The nodes represent the people in the network, and arcs represent the connections between them. The participants we consider are the traffickers, so-called ``bottoms", and victims. The term ``bottom" is used in the literature and in some operations to refer to someone who is a victim, but has gained trust and responsibility from the trafficker and thus has additional responsibilities within the trafficking operation \citep{belles2018defining}. They might be viewed as the highest ranking victim and function as a sort of right-hand person in the operation \citep{dank2014estimating}. However, there is a gray area in that bottoms may experience varying degrees of force and coercion since there is always a chance they will lose this status and be forced to trade sex \citep{roe2015sexual}. Likewise, a trafficker can expect or force some bottoms to exploit other victims \citep{roe2019six, belles2018defining}. For purposes of the generator, victims are split into two groups based on age, either minor or adult. Published research and insights from our survivor-centered advisory group suggest that there are operational differences in how traffickers interact with minors versus adults due to developmental differences and harsher legal penalties from trafficking minors \citep{marcus2014conflict}. 

In order to clarify the distinction between the activities of a specific trafficker and the activities of all traffickers, we additionally define the distinction between an operation and network. Henceforth, when we refer to an operation, we are referring to a single trafficker, their bottom (if they have one), and their victims. The network refers to all operations generated, where the number of operations is a user-specified input. We also want to provide a distinction to connections that are necessary for the function of a trafficking operations, as opposed to connections that are purely social, although we recognize that social connections may help further the activities of a trafficking operation. We define an operational arc as an arc that is necessary for the functions of the trafficking operation. An example of this is an arc between a trafficker and a victim, as that connection represents that the trafficker is able to control the victim. We define a social arc as an arc that does not necessarily have any immediate operational function associated with it. An example of this is an arc between two victims. Such a connection may not be necessary for the act of sex trafficking. However, if one of those victim were to be promoted to be the bottom, that arc would gain an operational function.

\subsection{\emph{Transdisciplinary Research Approach: Creation and Validation}}
\label{ssec:approach}

We apply a community-based participatory research approach in creating our network generator. Community-based participatory research approaches have been successfully applied to solve problems in health sciences and are especially impactful when the collaboration between researchers and community stakeholders is four years or longer \citep{brush2020success, haapanen2021community}. The survivor-centered advisory group brings together lived experience and deep social service and advocacy knowledge about sex trafficking, with many members possessing multiple forms of wisdom. The advisory group currently consists of five members, and two past members also contributed to this work. Together, advisory group members have over 90 years of combined experience in survivor-led social service, healing, and advocacy. For more details regarding the advisory group, their role, and transdisciplinary team building, see \cite{martin2022comms}. Our survivor-centered advisory group was formed in 2018 and has helped co-create research assumptions, approaches, knowledge, and research directions.  We then applied traditional, rigorous qualitative research methods, alongside operations research mathematical modeling, to inform this co-creation process.

Our research methods helped to synthesize and triangulate multiple sources of data to create and validate the assumptions built into the network generator. According to \cite{denzin2012}, p. 82, ``[t]he combination of multiple methodological practices, empirical materials, perspectives, and observers in a single study is best understood as a strategy that adds rigor, breadth, complexity, richness, and depth to any inquiry.”  Our transdisciplinary method inserts qualitative data at key decision-points in the mathematical modeling procedures, similar to the approach discussed by \cite{sharkey2021better}.  In particular, their approach discusses the importance of qualitative input into modeling decisions (which we have incorporated later in our novel conceptualization of flow in sex trafficking networks) and gathering data to populate resulting models.

For this first iteration of the network generator we used three sources of data: (1) documents from US federal prosecution of trafficking in Minnesota that were made publicly available (N=13); (2) targeted key informant interviews (N=10); and (3) secondary analysis of key informant interviews from previous studies (N=246) \citep{martin2017mapping, martin2014mapping}. From these sources, (1) is an example of agency-collected data, and (2) and (3) are examples of hidden data. In addition, our survivor-centered advisory group provided initial drawings of sample networks based on their experiences, and iteratively reviewed and critiqued assumptions and parameters in the network generator to assess whether it reflected their experiences and knowledge of other trafficking operations. Figure \ref{fig:AGdraw} in Appendix \ref{app:AGfigs} presents sample network drawings provided by the survivor-centered advisory group during one of our meetings with them. Data sources and analysis for this study focused on sex trafficking operations in the upper-Midwest region of the USA. For this first phase of development of the network generator, data analysis focused on smaller operations with only one trafficker. We used an analytic method called triangulation to identify and fill gaps within each data source, by comparing and contrasting content within and between sources.

\noindent \textbf{Analysis of Federal Prosecutions of Sex Trafficking Cases}: Research staff obtained the court case docket and indictment for cases prosecuted in Minnesota between 2009 and 2015. To transform this data into usable information for mathematical modeling, we extracted nodes (people, places, and things) and arcs (connections between nodes) and entered them into an excel spreadsheet. These data were then converted into network structure outputs for visualization. 

\begin{figure}[ht]
    \centering
    \includegraphics[width=0.55\linewidth]{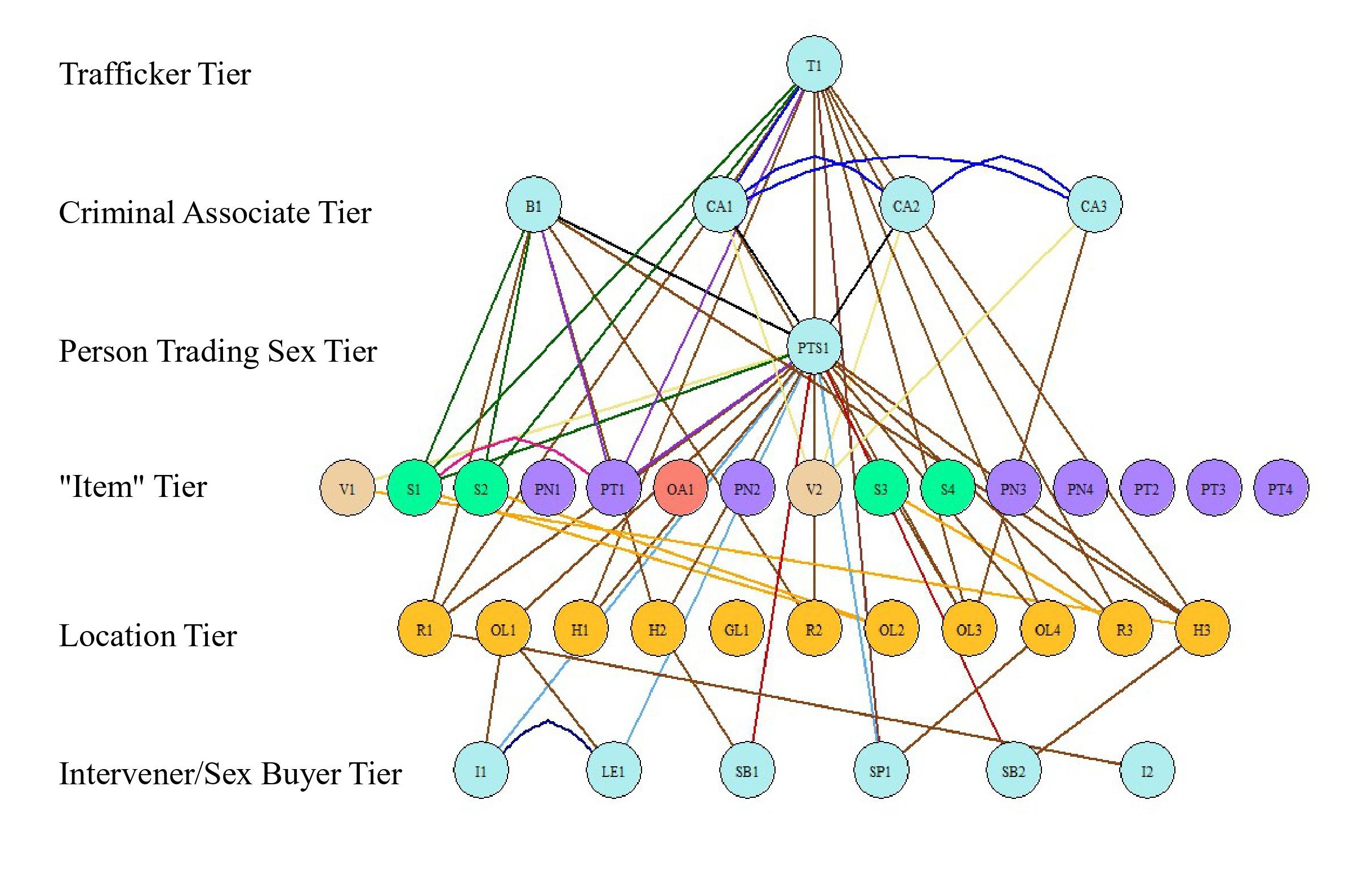}
    \caption{Sample network from one federal case file}
    \label{fig:sampleCFA}
\end{figure}

The research team, including the survivor-centered advisory group, visually inspected each output for verification and completeness. We wanted to know if these visualizations reflected what we would expect to see based on real-world experiences. Figure \ref{fig:sampleCFA}, provides an example of an output, visualizing the connections between people, places and things described in one of the federal case files. For this case, the advisory group noted that many key players and connections they would have expected to see where not represented in this diagram. Most notably, there was no record in the case file connecting the trafficker (T1) with their bottom (B1), which would be an expected connection. This direct connection may not have been necessary to prove the elements of trafficking in this case. The team identified many such instances where key relationships for the functioning of a trafficking network were not reflected in the case file documents. Across cases, we noted a lack of documentation of similarly important relationships within networks. Moreover, these case files spanned a short duration of time, thus only capturing a small snapshot of the trafficker’s operation in that moment in time. For the network from Figure \ref{fig:sampleCFA}, the recorded operations spanned one week.

\noindent \textbf{Analysis of Interview Data}: In semi-structured interviews with key informants, we asked 10 knowledgeable domain experts to provide details about the size and structure of relationship networks in trafficking operations, including number of victims and the social connections between victims. This data was thematically coded to extract typical ranges in size and organization structure of networks. Using this thematic framework, we systematically reviewed $260$ stakeholder interview data from past projects to extract additional information to guide model parameters and assumptions. We found agreement and coherence between the information garnered from key informant interviews and the secondary analysis.  

\noindent \textbf{Data Triangulation}: The above sources all have strengths and missing information. By triangulating these sources, combined with validation from the survivor-center advisory group, we developed a basis for the rules and parameters for network generation. We devised rules from consistencies across these multiple data sources, such as the distribution of number of victims within an operation, and used statistics and qualitative descriptions to estimate parameters for these rules. Once all necessary parameters were determined, we generated a set of networks to review with the qualitative researchers and survivor-centered advisory group to bring domain expertise in conversation with rules contained in the network generator. The OR team prepared presentations for the qualitative research team and advisory group to discuss the merits and shortcomings of the current iteration of the generator. This iterative process sought to validate the model against real world experience and expectation. Members of the qualitative research team produced detailed analytic memos and near verbatim transcriptions of survivor-centered advisory group meetings to capture critiques and missing information. These conversations resulted in adjustments to model parameters. Between meetings, questionnaires and surveys were prepared to help the OR team better implement the solicited feedback into the generator. This process of reviewing networks produced by the generator with the domain experts was repeated multiple times. 

We made several important changes to the network generator during each iteration. For example, the probability distribution of the number of victims in a trafficking operation was substantially revised based on advisory group feedback. Figure \ref{fig:hist} in Appendix \ref{app:AGfigs} displays the distribution incorporated in the current iteration of the generator. Discussions with the advisory group, in combination with data surfaced in the secondary review, also led us to revise the procedure the generator used to determine social connections between victims. We included the option for a trafficking network to separate victims into separate clusters who reside together, but do not necessarily know each other. The domain experts were then able to further refine the choice of parameters for this method, such as average and maximum cluster size. To the best of our knowledge, very little literature explores social connections between victims \citep{cockbain2018offender}. Conversations with the advisory group shed some light on these connections, such as smaller groups of victims being housed together and separated from the rest of the victims.

\subsection{\emph{Details of the Network Generator}}
\label{ssc:details}
Our network generator produces a network of $n_T$ single trafficker operations, where $n_T$ is an input. Interested researchers may access the network generator from the authors upon reasonable request. For each operation $i$, we first generate the number of victims $n_{vi}$ a trafficker has in their operation. The distribution we sample from is based on statistics gathered and validated by the domain experts.  Based on the number of victims, we randomly determine if there is a bottom in the network. The probability of a bottom in the network grows significantly with respect to the number of victims, with a bottom almost surely present when there are at least six victims. If it is determined there is a bottom in the network, a new node is added to be the bottom, as the statistics collected for number of victims did not include a bottom as a victim. These observations were made based on our secondary analysis of previously collected interview data in \cite{martin2014mapping, martin2017mapping}.

We next determine how a trafficker is managing the victims within their operation. We do so by partitioning the victims into ``clusters” \citep{melander2022conceptualizing}. Clusters can be thought of as groups of victims who were recruited roughly at the same time, or live in the same location. Each victim in a given cluster is connected to every other victim in the same cluster, i.e., a cluster forms a clique in the operational network. We first generate all feasible partitions of the number of victims, then remove partitions where any part is larger than six. This was determined to be the maximum size of a cluster based on our 
data sources and verified by our advisory group. For example, if there were seven victims in an operation, all partitions would be feasible except the case where all victims are in a single part. We then randomly select from the set of remaining partitions \citep{melander2022conceptualizing}. We estimated sizes of clusters, and distribution of cluster sizes $w$, by triangulating our multiple data sources and verified these parameters with our expert survivor-centered advisory group through group discussions and member surveys.  Let $\mathcal{P}$ be the set of all feasible clusters. We say $P \in \mathcal{P}$ is a set of clusters, where $P = (p_1, \ldots, p_{|P|})$, and $p_i$ is the number of victims in the $i^{th}$ cluster. When there is more than one cluster, we randomly determine the age group of the victims for the entire cluster, where the probability of the cluster consisting of minors increases with cluster size. When a cluster consists of two victims, we additionally allow for one victim to be a minor and one victim to be an adult. This is representative of a situation where a pair of victims may be family members (e.g. mother and daughter) or where there is a parental-type relationship. These choices regarding ages and clusters were suggested by the advisory group. 

We then determine how the trafficker (and bottom) interact with the clusters. If there is a bottom, we first determine how the victims are connected to the trafficker and bottom, as all victims are not necessarily connected to both the trafficker and the bottom. We randomly determine which clusters are connected to the trafficker, then which clusters are connected to the bottom. Similarly, if the trafficker (or bottom) is not randomly assigned any clusters, they will be assigned to the largest cluster. The probability of a cluster being assigned to the trafficker and bottom varies based on the age of the victims in the cluster and the size of the cluster. This is due to a trafficker potentially not wanting to have direct contact with a victim who is a minor, as prosecuting a sex trafficking case is easier if the victim is a minor, and the minimum punishments are more severe \citep{marcus2014conflict}. If there is no bottom, all victims are connected to the trafficker. If there is a single cluster, then age and connections to trafficker and bottom are determined individually, as if each victim were in their own cluster. We lastly refine upon the social network amongst victims within the operation by expanding upon the initial set of arcs provided by the clusters. We then determine any social connections between victims in different clusters. For each pair of victims in different clusters, we randomly determine if an arc between them should be added to the network. The probability of an arc being added between them is dependent on the age of the victims, since minor victims tend to be recruited into trafficking via their social network \citep{marcus2014conflict}. Algorithm \ref{alg:genOp} in Appendix \ref{app:formalAlg} formalizes the procedure of generating a single trafficker operation.

Each choice in the development of the generator was carefully made based on information from previous literature, the data sources reported in Section \ref{ssec:approach}, and meetings with the qualitative research team and the survivor-centered advisory group. Table \ref{tab:algstep} reports the main data source used to refine parameters used in the relevant steps in the generator. While the main sources were used to inform the initial choices, other sources were used to verify and refine the choices.

\begin{table}[]\footnotesize
    \centering
    \caption{Main data sources that inform each step in the network generation procedure}
    \begin{tabular}{|c|c|}
        \hline
        Step in Generator & Main Data Source \\
        \hline
        Number of Victims & Secondary Analysis of Interviews from Previous Studies \\
        \hline
        Inclusion of a Bottom & Targeted Interviews with Domain Experts \\
        \hline
        Age of Victims & Survivor-Centered Advisory Group \\
        \hline
        Social Arcs between Victims & Survivor-Centered Advisory Group \\
        \hline
    \end{tabular}
    \label{tab:algstep}
\end{table}

This procedure is repeated for the number of operations desired. After all operations are generated, we generate the social network amongst traffickers. From \cite{dank2014estimating}, we know that traffickers are connected to share information about profitable locations and law enforcement activity. Since little further is known about how traffickers are connected socially, we use the Watts-Strogatz model to generate the social network amongst traffickers \citep{watts1998collective}. The Watts-Strogatz model is often used for social networks since it produces networks with the “small world phenomena,” which indicates that the shortest paths between pairs of nodes include a small number of arcs and that nodes tend to be grouped into clusters, with a larger number of arcs between nodes in the same cluster than between nodes in different clusters. After the trafficker social network is generated, we then generate social connections between victims in different trafficking operations. Again, we impose a higher likelihood of two victims who are minors being connected over adult victims. Parameters regarding social connections were validated by the domain experts and advisory group. We formalize our generator in Algorithm \ref{alg:generator} in Appendix \ref{app:formalAlg}.

\section{Outputs}
\label{sec:output}
\begin{figure}[h]
    \centering
    \includegraphics[width=.75\linewidth]{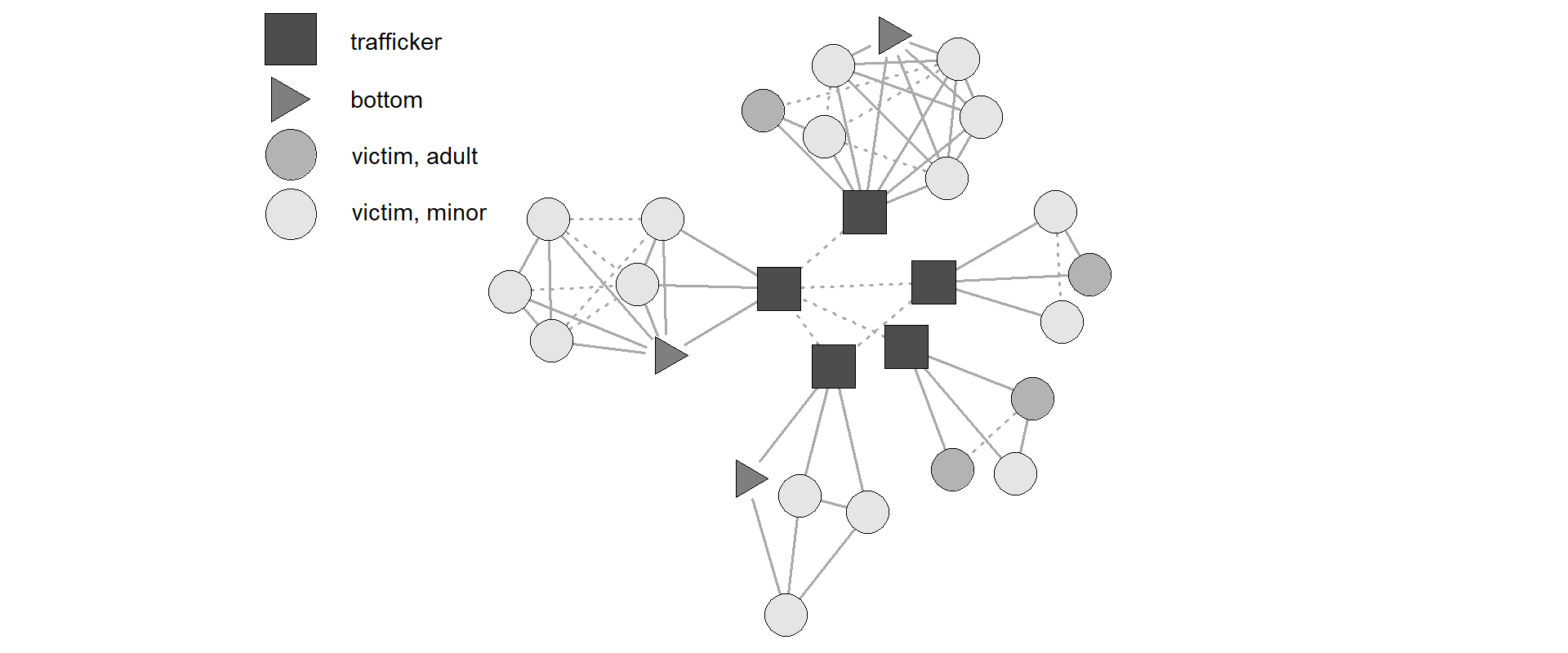}
    \caption{Sample sex trafficking network produced by the network generator}
    \label{fig:sampleNet}
\end{figure}

We now compare sample output operations from our domestic sex trafficking network generator to sex trafficking networks constructed in previous research.  Trafficker nodes are squares, bottom nodes are triangles, and victims nodes are circles. Nodes representing victims who are adults are a darker shade of gray than nodes representing victims who are minors. Arcs are solid if they are representing an operational arc (e.g. trafficker to victim), or dashed if they are representing a social arc. We include arcs between victims in the same cluster as an operational arc because we expect that victims in the same cluster will be living and working together. Figure \ref{fig:sampleNet} displays a sex trafficking network with five operations. Each operation is displayed individually in Appendix \ref{app:fig}.

In these five operations, three operations have a bottom. In $25$ generated outputs, $20$ operations have bottoms, and every operation with more than $5$ victims had a bottom. Of the eight operations with $4$ victims, only one operation does not have a bottom. Of the four operations with $3$, half of them have a bottom. Both networks with $2$ victims do not have a bottom. Additionally, of the $25$ operations, $5$ have one cluster, $16$ have two clusters, $2$ have three clusters, and $1$ has four clusters. In the operation with four clusters, three of them are isolated victims. The median cluster size has $2$ victims, with a maximum cluster size of $5$ victims. 

We compare centrality measures of the output operations in Figure \ref{fig:sampleOps} against those listed in \cite{cockbain2018offender} and the sample operations provided by the advisory group in Figure \ref{fig:AGdraw} (labeled AG). The advisory group provided two networks, a single trafficker operation, and a multiple trafficker operation with three traffickers. We consider each trafficker's operation in the multiple trafficker operation separately, to provide a more accurate comparison. We exclude Operation Retriever, as the number of victims in that network is significantly higher than the other networks in the text. Note that this analysis is comparing operations from two different locations (Midwest USA versus UK), and so trafficking networks may be structured differently between the two locations.

\begin{table}[h]\footnotesize
    \centering
    \caption{Comparison of centrality measures on victim networks between operations produced by the generator, provided by the advisory group, and the operations studied in \cite{cockbain2018offender}}
    \begin{tabular}{|c|c|c|c|c|c|}
        \hline
        Operation & Number of & Arc & Degree & Betweenness\\
        & Victims & Density & Centralization & Centralization \\
        \hline\hline
        1 & 3 & 0.667 & 0.333 & 1\\
        \hline
        2 & 3 & 0.667 & 0.333 & 1\\
        \hline
        3 & 6 & 0.733 & 0.267 & 0.200\\
        \hline
        4 & 3 & 1 & 0 & 0 \\
        \hline
        5 & 5 & 0.900 & 0.100 & 0.028\\
        \hline\hline
        AG1 & 2 & 1 & n/a\tablefootnote{Centrality scores can only be calculated in networks with at least three nodes.} & n/a \\
        \hline 
        %AG2 & 11 & 0.091 & 0.109 & 0.102 \\
        %\hline 
        AG2.1 & 5 & 0 & 0 & 0 \\
        \hline 
        AG2.2 & 3 & 0.667 & 0.333 & 1 \\
        \hline 
        AG2.3 & 4 & 0.333 & 0.333 & 0.333 \\
        \hline\hline
        Engage & 2 & 1 & n/a & n/a\\
        \hline
        Central & 4 & 1 & 0.0 & 0.0 \\
        \hline
        Span & 5 & 0.500 & 0.400 & 0.200 \\
        \hline
        Chalice & 6 & 0.400 & 0.600 & 0.300 \\
        \hline
    \end{tabular}
    \label{tab:sna}
\end{table}

The number of victims in operations produced by the network generator is similar to the number of victims in the operations investigated by Cockbain and in the operations described by the survivor-centered advisory group. Additionally, the arc densities and degree centralization scores of the synthetic operations all fall in the range of arc densities and degree centralization scores of the real operations. Excluding operation 4, operation AG1, operation Engage, and operation Central, which are all complete networks, the arc densities of the synthetic networks are higher than that of operations Span and Chalice, while the degree centralization scores of the synthetic networks are lower than the real networks investigated by Cockbain. However, the synthetic networks have degree centralization scores similar to that of the networks provided by the advisory group. This is likely due to the cluster structure in the synthetic networks, which was recommended by the advisory group. In operations with larger clusters, the arc density will be larger and the degree centralization will be smaller. The cluster structure also likely explains the significant difference in the betweenness centralization scores between the synthetic and real networks. While operation $3$ has a betweenness centralization score similar to operations Span and Chalice and AG2.3, operations $1$ and $2$ have betweenness centralization scores of $1$, matching AG2.2. This is because these three networks have the same structure: one cluster with two victims, and one isolated victim, with a single arc from the isolated victim to one victim in the cluster. These results suggest that although there is variation in the synthetic networks, they do share similarities with those produced through case file analysis on sex trafficking networks, as well as the networks provided by the advisory group. This is important as the generator is able to create more synthetic networks that are reflective of multiple data sources, which can be used for OR analysis.

We additionally provide an analysis of the spectrum of the Laplacian matrices of the networks to better understand the similarities in structures \citep{tantardini2019comparing}. Using spectral methods to compare networks was originally proposed by \cite{wilson2008study}, where they identify that the spectral distance between networks, i.e., the Euclidean distance between the eigenvalues of the Laplacian matrices, provides a strong measure of network similarity. For brevity, the tables of results appear in Appendix \ref{app:spec}. Table \ref{tab:spec1} compares the spectral distances between our synthetic networks and both the networks provided by our advisory group and the networks described in \cite{cockbain2018offender}. Table \ref{tab:spec2} additionally compares the spectral distances between advisory group and the networks described in Cockbain to provide an understanding of the deviation between real networks. In order to provide a comparison between networks of different sizes, we follow the procedure of \cite{wilson2008study} of adding zeros to the spectrum of the smaller network until the number of eigenvalues is the same. Since the spectral distance is a symmetric measure, we only include the upper right triangle for repeated entries in the table. We note that the largest eigenvalue from all networks is $10.207$ and the smallest eigenvalue is $0$.

Table \ref{tab:spec2} demonstrates that there can be a significant deviance in the victim networks in sex trafficking operations. In particular, Operation Retriever is significantly different from both the networks provided by the advisory group and the synthetic operations. This is likely due to the size differences between this network and other networks. We note that each data source seems to be more consistent with other networks from the same data source than the other data source. Table \ref{tab:spec1} demonstrates that our generator is able to produce networks similar to both data sources. In particular, Operations $1$, $2$ and $4$ are more similar to the networks provided by the advisory group, while Operations $3$ and $5$ are more similar to the networks shown in \cite{cockbain2018offender}.

\section{Responsible Use Case Study: Network Interdiction}
\label{sec:caseStudy}
We now present a case study on how an output of the network generator can be used as data for OR tools. We consider interdiction prescriptions via the max flow network interdiction problem (MFNIP) \citep{wood1993deterministic}. In a network $G = (N, A)$, with source node $s \in N$ and sink node $t \in N\setminus \{s\}$, and capacities on the arcs $u: A \rightarrow \mathbb{R}_{\ge 0}$, the max flow problem seeks to find the total amount of flow from $s$ to $t$ such that the flow on each arc is at most the capacity of that arc, and the amount of flow into a node is the same as the amount of flow leaving that node. The max flow network interdiction problem converts that problem to a two player game, where one player, known as the attacker, seeks to minimize the maximum flow through the network by choosing a subset of arcs to reduce their capacity to zero, subject to a budget and other constraints. The other player, the defender, then operates the network as per the max flow problem. Max flow network interdiction has successfully been applied to disrupting illicit drug trafficking networks \citep{baycik2018interdicting, malaviya2012multi, shen2021interdicting, kosmas2020interdicting}, and has been identified as an analytical tool to help address sex trafficking \citep{smith2020survey}. Some max flow network interdiction models have been proposed for disrupting human trafficking \citep{mayorga2019countering, tezcan2020human}. We apply the model of \cite{kosmas2020interdicting}, max flow network interdiction with restructuring (MFNIP-R), to networks outputted by our network generator. Their model accounts for how drug traffickers will respond to disruption efforts. This response is also a key aspect for disrupting sex trafficking, not merely displacing it. Accounting for responses from traffickers is vital to \emph{responsible} policy recommendations, as research suggests that removing individual victims from a trafficking situation, while clearly necessary, might paradoxically result in more victims being recruited into trafficking after an interdiction \citep{caulkins2019call, martin2014benefit}. This paradox aligns with the theoretical analysis in \cite{kosmas2020interdicting}, where they observed that current law enforcement policy typically recommended interdicting participants in drug smuggling networks that would trigger restructuring arcs that would cross the minimum cut.  We find that a similar situation occurs for our conceptualizations of sex trafficking networks, indicating that both disrupting current operations and their ability to restructure (especially recruit) is critical. 

\cite{kosmas2020interdicting} defines MFNIP-R as the following. Given a network $G = (N,A)$ with source node $s$, sink node $t$, set of restructurable arcs $A^R$, and node and arc capacities $u: N \cup A \cup A^R \rightarrow \mathbb{R}_{\ge 0}$, let $Y$ be the set of feasible interdiction plans and let $Z(y)$ be the set of feasible restructurings dependent on the chosen interdiction plan. First, the attacker chooses an interdiction plan $y \in Y$, setting the capacity of the interdicted nodes to $0$. Next, the defender chooses a restructuring plan $z \in Z(y)$ to add arcs to the network, increasing their capacity from $0$ to pre-specified values $u_z$. Lastly, the defender operates the network to maximize flow through the network. We note that, contrary to standard max flow network interdiction models, the model of \cite{kosmas2020interdicting} interdicts nodes instead of arcs, representing the removal of a participant in the network, rather than the connections between participants.  An equivalency between node interdiction and arc interdiction has previously been demonstrated in \citep{malaviya2012multi}.

\subsection{\emph{Modeling Domestic Sex Trafficking with Max Flow}}
\label{ssec:model}
Compared to drug trafficking, it is less clear in sex trafficking networks of traffickers and victims what an interdiction might seek to restrict in terms of flow. %The function of a drug trafficking network is to move and eventually sell a product (i.e. drugs) to a user. Thus the flow is the drugs. In a sex trafficking network, flow is more complex. 
The ``product" being sold in sex trafficking is people and a sexual experience \citep{martin2017mapping, martin2014mapping}. The commercialization and sale of sex to a sex buyer may in some ways be analogous to the sale of drugs since both are commodities. Yet, the flow within the network is not the same because people (and their labor) are not equivalent to drugs. Much more work is needed to conceptualize how this concept of ``flow" should best be applied to sex trafficking networks. Thus, we posit a novel interpretation of the maximum flow problem as it pertains to traffickers and victims of trafficking within a sex trafficking operation. Flow could represent the ways traffickers exhibit control over victims and the sex acts that victims are forced to perform. Thus flow is exerted through control over trafficking victims in order to make them (and their labor) a product to sell \citep{martin2014mapping}. This conceptualization was developed in collaboration with the domain experts to reflect that the network flows in our approach are of different nature compared to traditional supply chain models.

In a network framework, we let the capacities of trafficker and bottom nodes represent the number of victims they are able to manage, and the capacities of victims represent that they are able to be controlled. Interdicting a node represents that node can no longer control or be controlled, and restructuring an arc represents establishing a connection that allows for a trafficker or bottom to control a victim. We note that, in our model, an interdicted node has its capacity set to zero, representing the participant being fully removed from the network. An interdiction can be practically interpreted as any action that removes a participant from the network, such as law enforcement arresting a trafficker.  Another example would be an agency providing a victim access to resources (e.g., safe and secure housing, money) that would allow them to be able to leave their trafficker.  It must be noted that the flow that is being modeled is exerted through violence, manipulation, and harm. For this discussion, we will focus on control within the network and on situations in which a trafficker needs to maintain control over a certain number of victims. For the sake of the model, we postulate that a trafficker has a specific number of victims they wish to have under their control at any given time.  This is a preliminary exploration; more empirical research is needed to verify and extend modeling to understand the specificity and nuances of sex trafficking.

\begin{figure}
\begin{subfigure}{.48\textwidth}
  \centering
  \includegraphics[width=.95\linewidth]{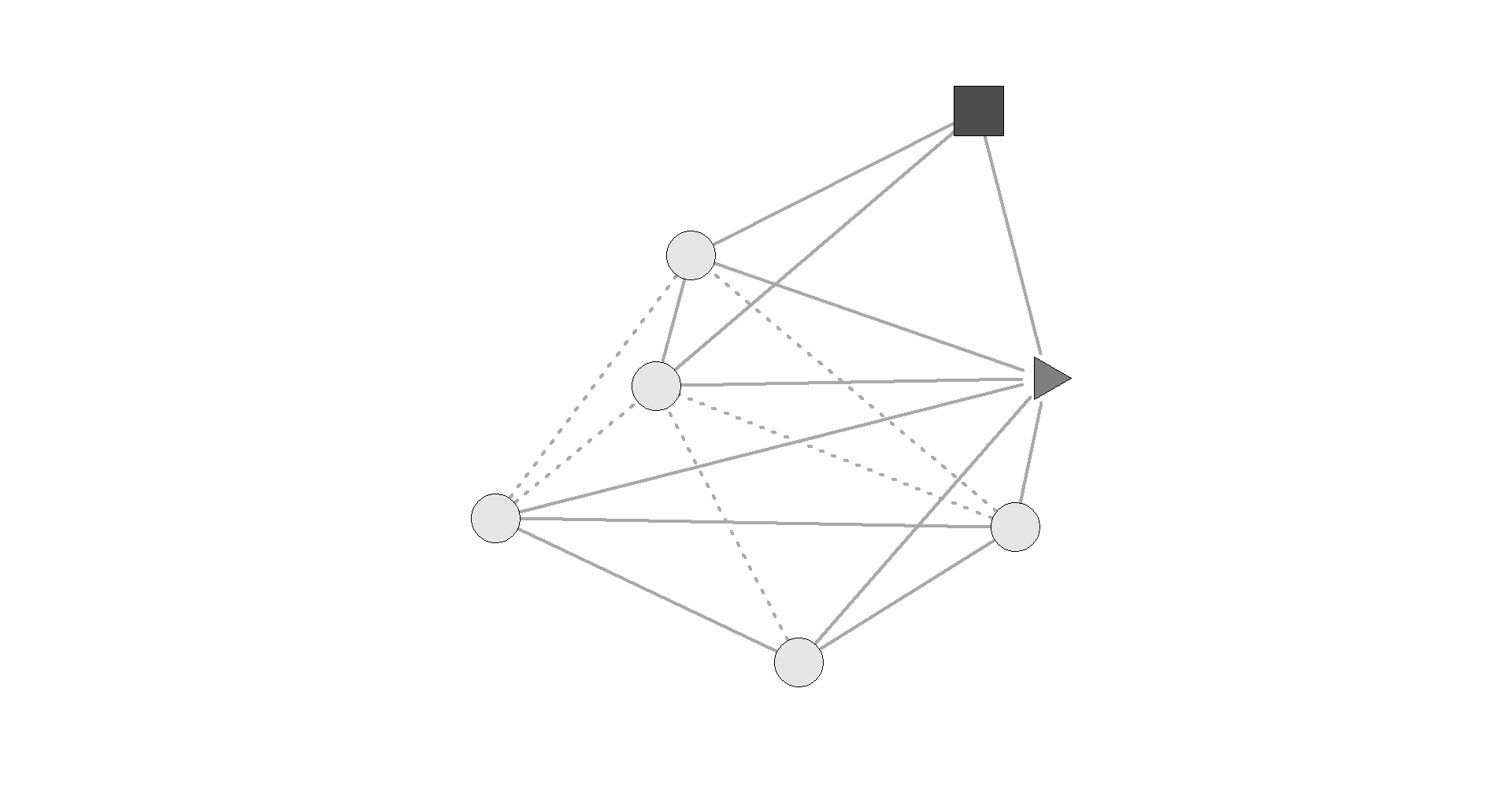}
  \caption{Operation produced by network generator}
  \label{fig:op5undirected}
\end{subfigure}%
\begin{subfigure}{.48\textwidth}
  \centering
  \includegraphics[width=.95\linewidth]{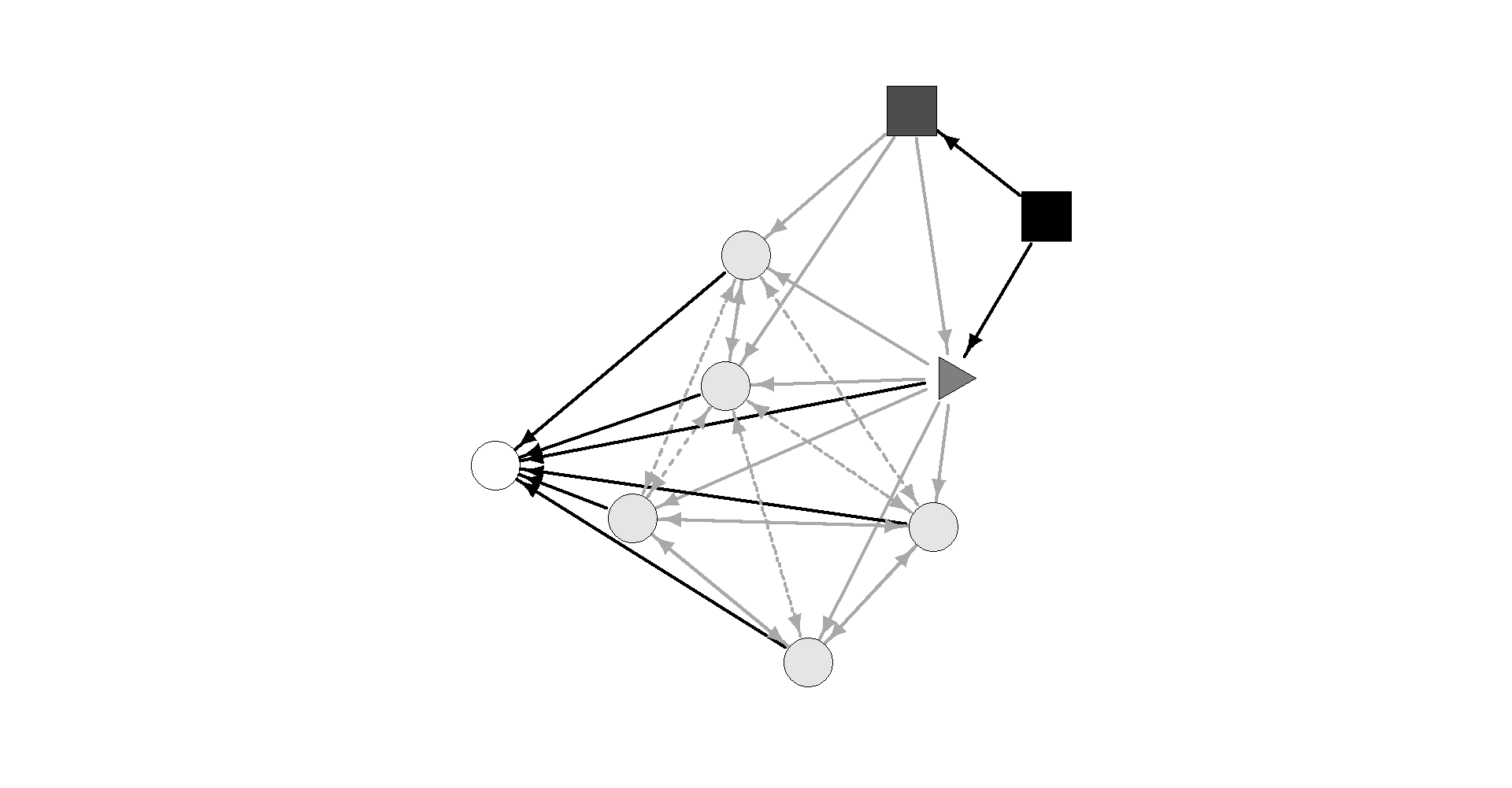}
  \caption{Operation usable in max flow setting}
  \label{fig:op5directed}
\end{subfigure}
\caption{Demonstration on converting operation $5$ to be usable in a max flow setting}
\label{fig:conversion}
\end{figure}

To convert an output of the network generator to a network usable in max flow network interdiction, we need to add in a source node and sink node. We add arcs from the source node to each trafficker and bottom node with infinite capacity, and arcs from each bottom and victim node to the sink with capacity one. All arcs incident to the trafficker are directed out of the trafficker, and all arcs incident to the bottom and a victim are directed out of the bottom and into the victim. Arcs between victims are replaced with a pair of arcs between the victims, one in each direction. The capacity of each trafficker node is the number of victims they can control, the capacity of each victim node is one, and the capacity of each bottom node is the number of victims they can control plus one. This is to recognize that bottoms are victims too, and that their role in sex trafficking networks does not make them any less of a victim than the other victims.  Note that, with our modeling choices, it is not necessary to directly model the control of a bottom with an arc from the trafficker to them as the arc from the source can model this control, even though this arc will always be included in operations produced by the network generator. For ease of notation, we define $T$ to be the set of traffickers, $B$ to be the set of bottoms, and $V$ to be the set of victims. Capacities for the traffickers and bottoms are chosen in such a way that the maximum flow through the network is the total number of victims currently in the network (including bottoms). The arcs between traffickers are only used to determine restructurable arcs, and are assigned capacity $0$. Figure \ref{fig:conversion} demonstrates this conversion on a sample operation, where the black square node is the source node, the white circle node is the sink node, and the black arcs are the newly added arcs.

We now define the set of restructurable arcs. The types of restructurable arcs were determined by analyzing transcripts from previous meetings with the qualitative research team and survivor-centered advisory group. For the types of restructurable arcs, we allow for traffickers to recruit each other's victims (or likewise, a victim choosing to join a different trafficker's operation), traffickers to give or take victims from their bottom, the recruitment of new victims, back-up traffickers to take over an interdicted trafficker’s operation, the promotion of a new bottom, and traffickers to give victims to their newly promoted bottom. As in \cite{kosmas2020interdicting}, we define sets of restructurable arcs $A^R = A^{R,out} \cup A^{R,in}$, based on whether a restructurable arc can be initiated by a trafficker or by a victim. An ``out" restructuring (arcs belonging to $A^{R,out}$) represents a trafficker restructuring to a new victim after having one of their victims interdicted. An ``in" restructuring (arcs belonging to $A^{R,in}$) represents a victim being recruited into a new operation after their trafficker has been interdicted. We start with $A^{R,out}$ and $A^{R,in}$ consisting of the set of arcs between traffickers and the victims of traffickers they know. For $i \in T$ and $k \in V$, $(i,k) \in A^{R,out}$ and $(i,k) \in A^{R,in}$ if there exists a $h \in T$ such that $(i,h) \in A$ and $(h,k) \in A$. To allow for a trafficker $i$ to assign a new victim $k$ to their bottom $j$, we include $(j,k) \in A^{R,out}$ if $(j,k) \notin A$. Likewise, we allow for a trafficker $i$ to take a victim $k$ from their bottom $j$ by including $(i,k) \in A^{R,out}$ if $(i,k) \notin A$ and $(j,k) \in A$.

To model the recruitment of new victims, we introduce an additional set of nodes $V^R$. We include an arc from each of these nodes to the sink node in the arc set $A$. We then randomly determine a subset of traffickers that can restructure to each recruitable victim node, and add arcs between the traffickers and the recruitable victim to $A^{R,out}$. This results in these nodes having no arcs into them in the initial network, meaning no flow can pass through them, but flow can pass through them if an arc ending at the recruitable victim is restructured by a trafficker. To model an operation having a back-up trafficker, we introduce an additional set of nodes $T^R$ to be the set of back-up traffickers for certain operations, which is represented by ordered pairs $(i,h)$, where trafficker $i$ can be replaced by back-up trafficker $h$ if trafficker $i$ is interdicted. We add arcs from the back-up trafficker $h$ to all of trafficker $i$'s victims to the arc set, and include an arc $(s,h)$ in the set of restructurable arcs.  As with recruitable victims, there are no arcs into back-up trafficker $h$ in the initial network, so no flow can pass through the node, but restructuring the arc $(s,h)$ allows for flow to pass through the node. To model the promotion of new bottoms, we define the set $B^R$ as the set of victims which may be eligible to be promoted to the role of bottom, which is represented by ordered pairs $(j,k)$, where victim $k$ may be promoted if bottom $j$ was interdicted. If victim $k$ is promoted to the role of bottom, their capacity is increased by $\Tilde{u}_k$, allowing flow to pass from victim $k$ to the victims that they are adjacent to. We include arc $(s,k)$ in the set of restructurable arcs $A^{R,out}$, indicating that if this arc is restructured, the newly promoted bottom is able to receive flow from the source as the original bottom did. We also allow for a trafficker $i$ to assign a victim $l$ to the newly promoted bottom $k$ by including arc $(k,l) \in A^{R,out}$ if $(k,l) \notin A$ that can only be restructured if $(s,k)$ has been restructured. The capacities of back-up traffickers are chosen to be $75\%$ of the capacity of the initial trafficker (minimum $1$), and the capacity increase of promoting a victim to be a bottom is chosen to be $66\%$ of the capacity of the initial bottom (minimum $1$). These choices were made to reflect a decrease in network functionality due to a new person taking over the act of trafficking.

Restructurable arcs between an existing trafficker and existing victim are included in $A^{R,in}$, while all restructurable arcs are included in $A^{R,out}$. Note that $A^{R,in} \subset A^{R,out}$ in our model. This is because the only arcs belonging to $A^{R,in}$ are between a victim currently in the network and a trafficker, which can also be restructured by the trafficker. Thus, these arcs also belong to $A^{R,out}$.

For brevity, we define the formal mathematical model in Appendix \ref{app:model}. We apply the method of \cite{kosmas2020interdicting} to solve this model. Their method is a column-and-constraint generation (C\&CG) method, where, when a previously visited restructuring plan is infeasible with respect to the current interdiction plan, the only feasible components in that restructuring plan are used to determine a lower bound on the true objective value. This method has been shown to be effective in solving their model faster than standard C\&CG methods, particularly when recruitment is modeled.

\subsection{\emph{Computational Results}}
\label{ssec:comp}
We implement this interdiction model on five generated networks, each with five operations, as well as a five operation network provided by the advisory group. The advisory group network was created by merging two copies of the single trafficker operation and the multiple trafficker operation to create a network with the same number of traffickers as the synthetic networks. Table \ref{tab:netstats} reports the number of nodes, and number of bottoms and victims, in each network.

\begin{table}[h]\footnotesize
    \centering
    \caption{Sizes of generated networks}
    \begin{tabular}{|c|c|c|c|c|}
        \hline
        Network & Number of Nodes & Number of Traffickers & Number of Bottoms & Number of Victims \\
        \hline
        AG & 25 & 5 & 5 & 15 \\
        \hline
         1 & 28 & 5 & 3 & 20 \\
         \hline
         2 & 32 & 5 & 4 & 23 \\
         \hline
         3 & 35 & 5 & 5 & 25 \\
         \hline
         4 & 35 & 5 & 4 & 26\\
         \hline
         5 & 27 & 5 & 4 & 18\\
         \hline
    \end{tabular}
    \label{tab:netstats}
\end{table}

We set the cost of interdicting a trafficker to be $8$, the cost to interdict a bottom to be $4$, and the cost to interdict a victim to be $2$. Interdicting a bottom reduces the cost of interdicting their trafficker by $3$, and interdicting a victim reduces the cost of interdicting their trafficker by $1$. Each trafficker has a budget of $8$ to restructure their operation. The cost to restructure to a victim currently in the network is $1$, as is the cost to take a victim from their bottom or give a victim to their bottom. The cost to recruit a new victim not currently in the network is $2$. The cost for a back-up trafficker to come into the operation is $4$. The cost to promote a new bottom is $5$, and the cost for the trafficker to give them a victim is $2$. We set the number of recruitable victims to be $40\%$ of the number of victims in the network. A trafficker will have a back-up trafficker if they have at least four victims (including a bottom). If a trafficker has a bottom, the number of victims that can be promoted to become a bottom is half of the number of victims the trafficker is connected to (minimum one). We note that parameters were chosen to demonstrate how a network interdiction model would be applied to these networks, and that future work in applying network interdiction to sex trafficking networks would require further collaboration and validation with domain experts.

\begin{figure}[h!]
\centering
\begin{subfigure}{.48\textwidth}
  \centering
  \includegraphics[width=.8\linewidth]{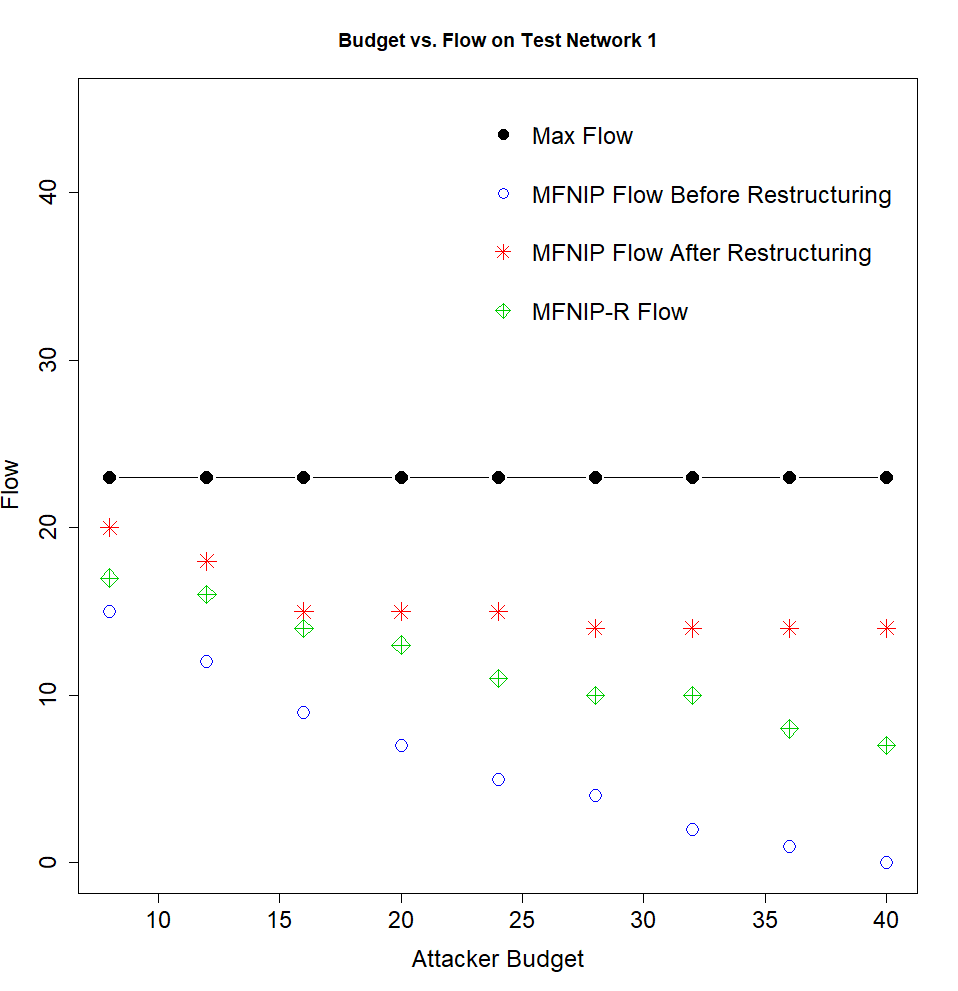}
  \label{fig:flow1}
\end{subfigure}%
\begin{subfigure}{.48\textwidth}
  \centering
  \includegraphics[width=.8\linewidth]{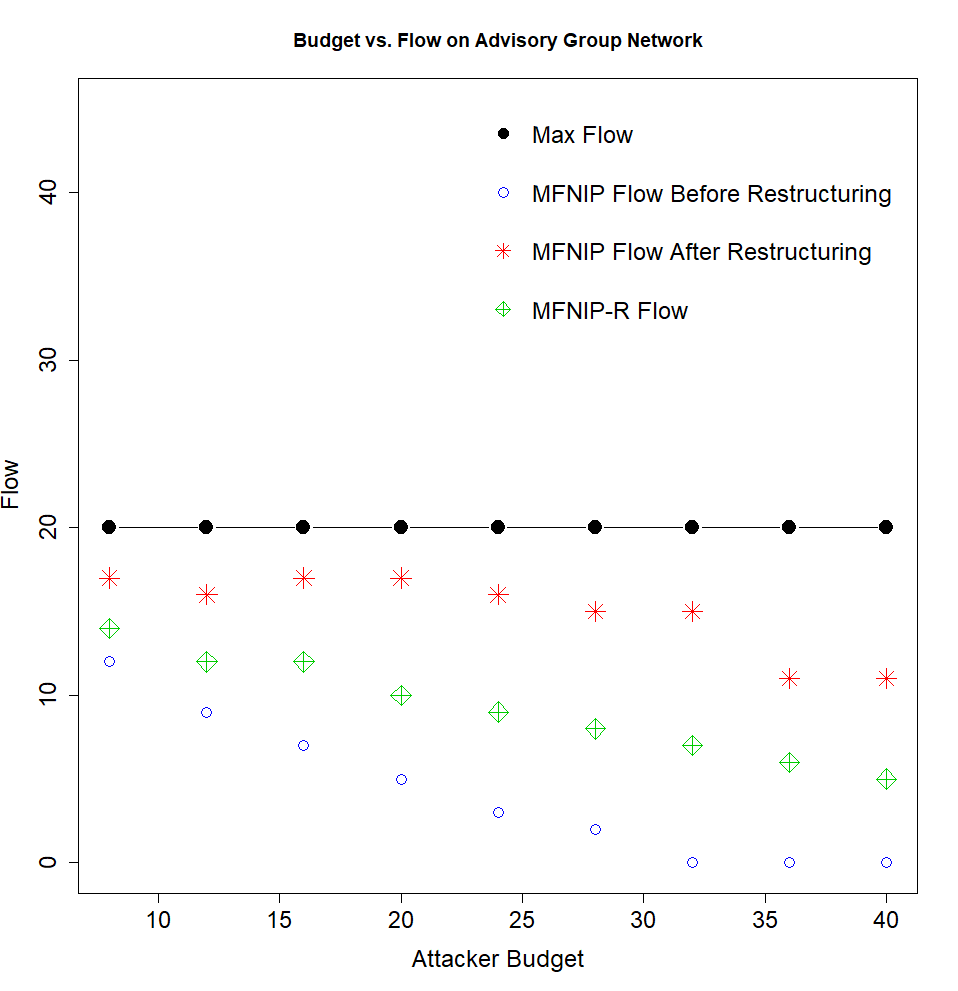}
  \label{fig:flowAG}
\end{subfigure}
\caption{MFNIP-R flows over varying attacker budgets}
\label{fig:flowplot}
\end{figure}

All instances were solved within $2$ hours. Experiments were conducted on a laptop with an Intel\textsuperscript{\textregistered} Core\textsuperscript{TM} i5-8250 CPU @ 1.6 GHz - 1.8 GHz and 16 GB RAM running Windows 10 using AMPL with Gurobi 9.0.2 as the solver. Figure \ref{fig:flowplot} demonstrates the flow through the network over different attacker budgets. Results on other networks can be found in Appendix \ref{app:moreResults}. In this figure, the dotted black represents the total number of victims (including bottoms) in the network. The hollow circles represent the interdicted flow determined by MFNIP. The stars represent the flow after optimally restructuring the network in response to MFNIP's interdictions. The diamonds represent the interdicted flow determined by MFNIP-R. Network $1$ is the network in Figure \ref{fig:sampleNet}.

In each network and across all budget levels, the traffickers are able to recover a significant amount of flow after restructuring in response to the MFNIP recommended interdictions. This is primarily due to being able to recruit new victims when victims are primarily interdicted, and being able to replace interdicted traffickers with back-up traffickers when traffickers are primarily interdicted. However, when we are able to account for how the traffickers will restructure in response to interdiction, we are often able to reduce the flow. Table \ref{tab:whoInt} demonstrates the interdiction recommendations for the network in Figure \ref{fig:sampleNet}. Columns $2$ - $4$ display the number of traffickers, bottoms, and victims that MFNIP recommends to interdict, and columns $5$ - $7$ display the number of traffickers, bottoms, and victims that MFNIP-R recommends to interdict.

\begin{table}[h]\footnotesize
    \centering
    \caption{Interdiction recommendations from MFNIP and MFNIP-R for Network $1$}
    \begin{tabular}{|c|c|c|c|c|c|c|}
          \hline
         Attacker & MFNIP & MFNIP & MFNIP  & MFNIP-R & MFNIP-R & MFNIP-R \\
         Budget & Int Trafficker & Int Bottom & Int Victim & Int Trafficker & Int Bottom & Int Victim \\
         \hline
         8 & 0 & 2 & 0  & 0 & 0 & 4 \\
         \hline
         12 & 1 & 1 & 0  & 1 & 0 & 4 \\
         \hline
         16 & 1 & 2 & 1  & 1 & 0 & 6 \\
         \hline
         20 & 1 & 2 & 3  & 1 & 1 & 6\\
         \hline
         24 & 1 & 3 & 3  & 1 & 0 & 10 \\
         \hline
         28 & 1 & 3 & 4  & 1 & 2 & 8 \\
         \hline
         32 & 1 & 3 & 8  & 2 & 1 & 8 \\
         \hline
         36 & 1 & 3 & 10 & 2 & 1 & 12 \\
         \hline
         40 & 1 & 3 & 12 & 2 & 3 & 10 \\
         \hline
    \end{tabular}
    \label{tab:whoInt}
\end{table}

\begin{table}[]\footnotesize
\centering
    \caption{Interdiction recommendations from MFNIP and MFNIP-R for Network provided by the survivor-centered advisory group}
\begin{tabular}{|c|c|c|c|c|c|c|}
\hline
Attacker & MFNIP & MFNIP & MFNIP & MFNIP-R & MFNIP-R & MFNIP-R \\
Budget & Int Trafficker & Int Bottom & Int Victim & Int Trafficker & Int Bottom & Int Victim \\ \hline
8 & 0 & 2 & 1 & 0 & 0 & 4 \\ \hline
12 & 0 & 4 & 0 & 0 & 0 & 6 \\ \hline
16 & 0 & 5 & 0 & 0 & 1 & 6 \\ \hline
20 & 0 & 5 & 2 & 0 & 2 & 6 \\ \hline
24 & 0 & 5 & 4 & 1 & 1 & 8 \\ \hline
28 & 1 & 5 & 2 & 1 & 2 & 9 \\ \hline
32 & 2 & 5 & 2 & 2 & 2 & 9 \\ \hline
36 & 0 & 5 & 10 & 2 & 3 & 9 \\ \hline
40 & 0 & 5 & 10 & 3 & 3 & 8 \\ \hline
\end{tabular}
\end{table}

A clear distinction between the two recommended plans is the number of bottoms that each model recommends to interdict. Consistently across all attacker budgets, MFNIP recommends interdicting at least as many bottoms than MFNIP-R does. This observation is consistent across all tested networks. This is likely due to MFNIP-R identifying that bottoms have limited ability to restructure the network, as the only way they can acquire new victims is if their trafficker assigns them new victims. This finding is also consistent with the network provided by the advisory group, where MFNIP recommends interdicting all bottoms as soon as the budget allows for it, and continues this recommendation for every larger budget level. This seems to highlight that limiting the ability of the trafficker to restructure after disruptions may be just as important as disrupting the current operations, which is similar to the findings of \cite{kosmas2020interdicting}. Additionally, network $3$, where every operation has a bottom, had the smallest gap in flows projected by MFNIP and MFNIP-R. These observations suggest that better understanding how bottoms help operate the network both before and after disruptions may be critical to understanding how to more effectively dismantle sex trafficking networks. The advisory group confirmed the importance of these findings in implementing effective disruptions.

\section{Limitations and Directions for Future Qualitative Research}
\label{sec:limit}
As the lack of data is the challenge we wish to address, our network generator has many limitations that a responsible user needs to be aware of. First and foremost, our network generator is built on sources about domestic sex trafficking cases within the Midwest of the USA, since the case file analysis focused on the cases in the Midwest and the advisory group’s experiences involve operations within the Midwest. As such, their experience is not necessarily representative of sex trafficking as a whole. It may be that biases based on their experience are built into the network generator. However, the more the network generator is used and feedback is provided regarding operations in more locations and contexts, the more accurate we will be able to make it over time. Users should collaborate with domain experts in their geographical location to understand how sex trafficking networks in their location may operate differently from that of the Midwest.

Another limitation is regarding the social connections between traffickers. It is known that some traffickers are connected socially with other traffickers, but there has been little research on what these social networks look like. This gap prevents us from developing more authentic trafficker social networks. Further research focused on how traffickers are connected, both socially and professionally, will allow for more accurate trafficker social networks to be produced.

How bottoms interact with the network is also an area where further research is needed. While the role of bottoms in their own trafficking operation has been previously studied, little is known on how they are connected to other traffickers and victims outside of their operation. It is also not well known what causes a trafficker to want or need to have a bottom in their network. Here, we base the likelihood of there being a bottom on the number of victims in the trafficker’s operations. Future qualitative research can ascertain why a trafficker chooses to promote a victim to the role of the bottom and how having a bottom allows a trafficker to grow their business.

\section{Conclusions}
\label{sec:conc}
We presented our network generator for producing synthetic domestic sex trafficking networks. The networks produced include the operational and social connections between traffickers, bottoms, and victims. Data sources including publicly available federal case files and interviews with domain experts were triangulated to determine the basis and parameters in the generator. The generator was further validated by domain experts, including a survivor-centered advisory group. As qualitative research furthers our understanding of sex trafficking networks, we can refine and advance the functionality of the generator in future iterations. An example of this is on how traffickers collaborate, which would allow us to build multiple-trafficker operations. The network generator and generated networks in this study are available from the corresponding author upon reasonable request.

This generator allows for the OR community to engage in human trafficking research without needing to go through the time-intensive process of collecting and cleaning their own data. We demonstrate how network interdiction can be applied to networks produced by this generator. In particular, we proposed a novel conceptualization of flow which considers the ability of the traffickers to control their victims. Our results suggest that better understanding the roles of bottoms in maintaining operations after disruptions occur may be key to more impactful disruptions.

The advisory group has suggested three avenues of future research to improve the applicability of the network generator. The first avenue is to include a temporal component. From their insights, the victims in an operation can change rapidly (in some cases daily) depending on many different factors. Being able to account for a temporal component can allow for a better understanding of how long term disruption decisions can be made. The second avenue of future research is to include a spatial component. Trafficking operations may move to different locations based on profit and law enforcement activity. Understanding how trafficking operations move between different locations will be essential to effective disruption prescriptions. The third avenue of future research is to produce networks where multiple traffickers are collaborating in larger trafficking operations.

\section*{Acknowledgements}
This material is based upon work supported by the National Science Foundation (NSF) under Grant No. 1838315 and the National Institute of Justice (NIJ) under Grant No. 2020-R2-CX-0022. The opinions expressed in the paper do not necessarily reflect the views of the NSF or NIJ. We recognize that our research cannot capture all the complexities of the lived experiences of trafficking victims and survivors.  We acknowledge the significant contributions of our survivor-centered advisory group including: Tonique Ayler, Breaking Free, Housing Advocate, Survivor Leader; Terry Forliti, Independent Consultant, Survivor Leader;  Joy Friedman, Independent Consultant, Survivor Leader; Mikki Mariotti, Director, The PRIDE Program of the Family Partnership; Christine Nelson, Independent Consultant, Fellow of Survivor Alliance, Survivor Leader; Lorena Nevile, Vice President of Programs, The Family Partnership; and Drea Sortillon, Witness-Victim Division, Hennepin County Attorney’s Office.  They have provided critical expertise in understanding the complexities of trafficking networks. We appreciate the recommendations of the reviewers, which have helped to better present the qualitative approaches used to determine parameters of the network generator.

\bibliographystyle{agsm}
\spacingset{1}
\bibliography{ref}

\newpage
\begin{appendices}
\section{Additional Figures}
\label{app:AGfigs}
\begin{figure}[h]
    \centering
    \begin{subfigure}{\textwidth}
        \centering
        \includegraphics[width=\linewidth]{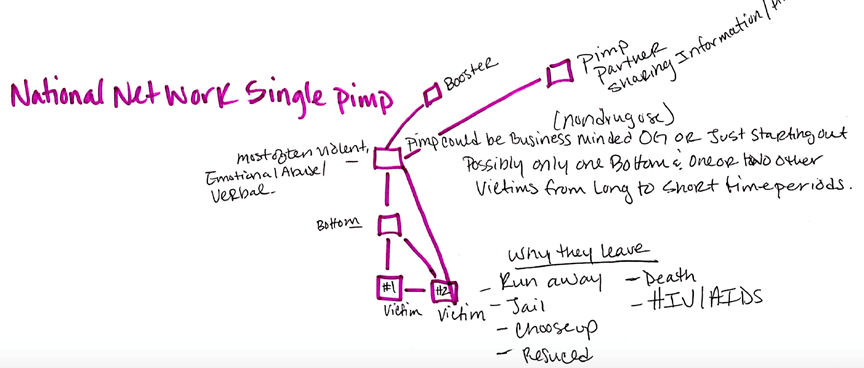}
        \caption{Single trafficker operation}
        \label{fig:ags}
    \end{subfigure} 
    
    \begin{subfigure}{\textwidth}
        \centering
        \includegraphics[width=\linewidth]{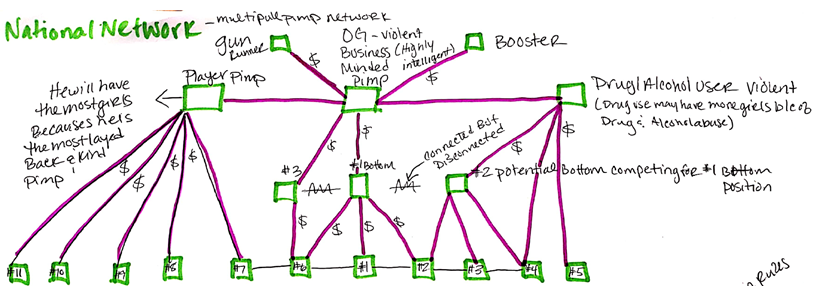}
        \caption{Multiple trafficker operation}
        \label{fig:agm}
        \end{subfigure}
    \caption{Sample network drawings provided by the survivor-centered advisory group}
    \label{fig:AGdraw}
\end{figure}

\begin{figure}[ht]
    \centering
    \includegraphics[width=0.55\linewidth]{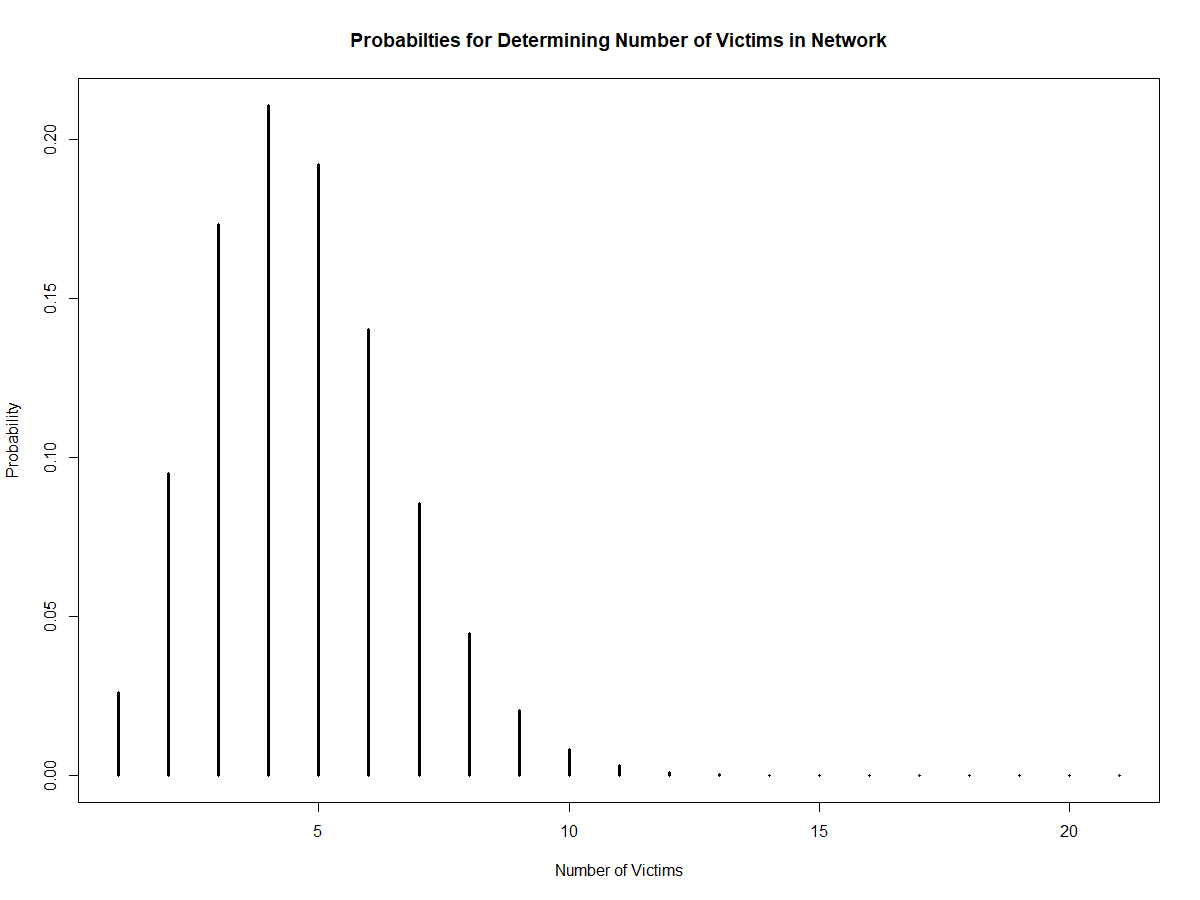}
    \caption{Probability distribution of number of victims in a trafficking operation, validated with domain experts}
    \label{fig:hist}
\end{figure}

\clearpage
\section{Formal Algorithms}
\label{app:formalAlg}
\begin{algorithm}\scriptsize
\caption{Generation Procedure for a Single Trafficker Operation ($GenerateOp(i)$)}
\label{alg:genOp}
\begin{algorithmic}
\STATE{\textbf{Initialize:}} operation number $i$.
\STATE{\textbf{Step 1.} Generate number of victims $n_{vi}$ for operation $i$.}
\STATE{\textbf{Step 2.} Randomly determine if operation $i$ has a bottom $b$.}
\IF{$n_{vi} > 1$}
\STATE{\textbf{Step 3a.} Generate set of feasible partitions $\mathcal{P}$ of $n_{vi}$.}
\STATE{\textbf{Step 3b.} Randomly select from $P \in \mathcal{P}$ using weighted probabilities $w_P$.}
\ELSE
\STATE{\textbf{Step 3c.} Set $P = \{1\}$}.
\ENDIF
\IF{$|P| \ge 2$}
\FOR {$j = 1:|P|$}
\STATE{\textbf{Step 4a.} Randomly generate age group of cluster of victims $p_j \in P$.}
\ENDFOR
\ELSE
\FOR{$j=1:n_{vi}$}
\STATE{\textbf{Step 4b.} Randomly generate age group of victim $v_j \in P$.}
\ENDFOR
\ENDIF

\IF{Operation $i$ has a bottom}
\IF{$|P| \ge 2$}
\FOR{$j=1:|P|$}
\STATE{\textbf{Step 5a.} Randomly determine if cluster of victims $p_j$ is directly connected to the trafficker, bottom, or both.}
\ENDFOR
\ELSE
\FOR{$j=1:n_{vi}$}
\STATE{\textbf{Step 5b.} Randomly determine if victim $v_j$ is directly connected to the trafficker, bottom, or both.}
\ENDFOR
\ENDIF
\ELSE
\FOR{$j=1:n_{vi}$}
\STATE{\textbf{Step 5c.} Connect victim $v_j$ to the trafficker.}
\ENDFOR
\ENDIF
\IF{$|P| \ge 2$}
\FOR{$j=1:(|P|-1)$}
\FOR{$k=(j+1):|P|$}
\FOR{$u \in p_j$}
\FOR{$v \in p_k$}
\STATE{\textbf{Step 6}. Randomly determine whether or not victims $u$ and $v$ are connected.}
\ENDFOR
\ENDFOR
\ENDFOR
\ENDFOR
\ENDIF
\end{algorithmic}
\end{algorithm}

\clearpage
\begin{algorithm}\scriptsize
\caption{Network Generator for Networks of Single Trafficker Operations}
\label{alg:generator}
\begin{algorithmic}
\STATE{\textbf{Initialize:}} number of traffickers $n_T$.
\FOR{$i = 1:n_T$}
\STATE{\textbf{Step 1.} Generate single trafficker operation $i$ with $GenerateOp(i)$.}
\ENDFOR
\STATE{\textbf{Step 2.} Generate trafficker social network.}
\end{algorithmic}
\end{algorithm}

\section{Sample Generated Operations}
\label{app:fig}
\begin{figure}[h!]
\centering
\begin{subfigure}{.48\textwidth}
  \centering
  \includegraphics[width=\linewidth]{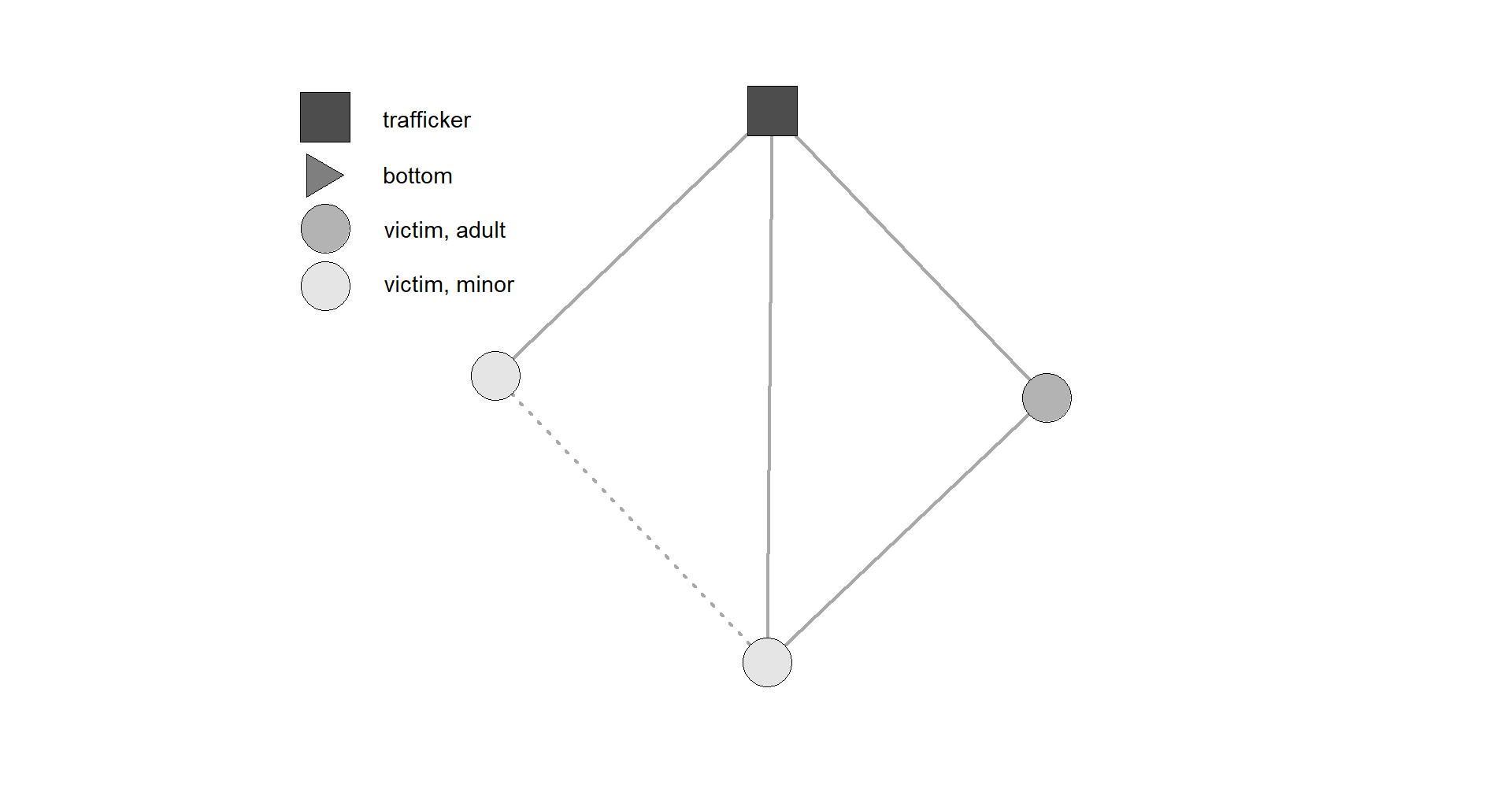}
  \caption{Sample operation 1}
  \label{fig:op1}
\end{subfigure}%
\begin{subfigure}{.48\textwidth}
  \centering
  \includegraphics[width=\linewidth]{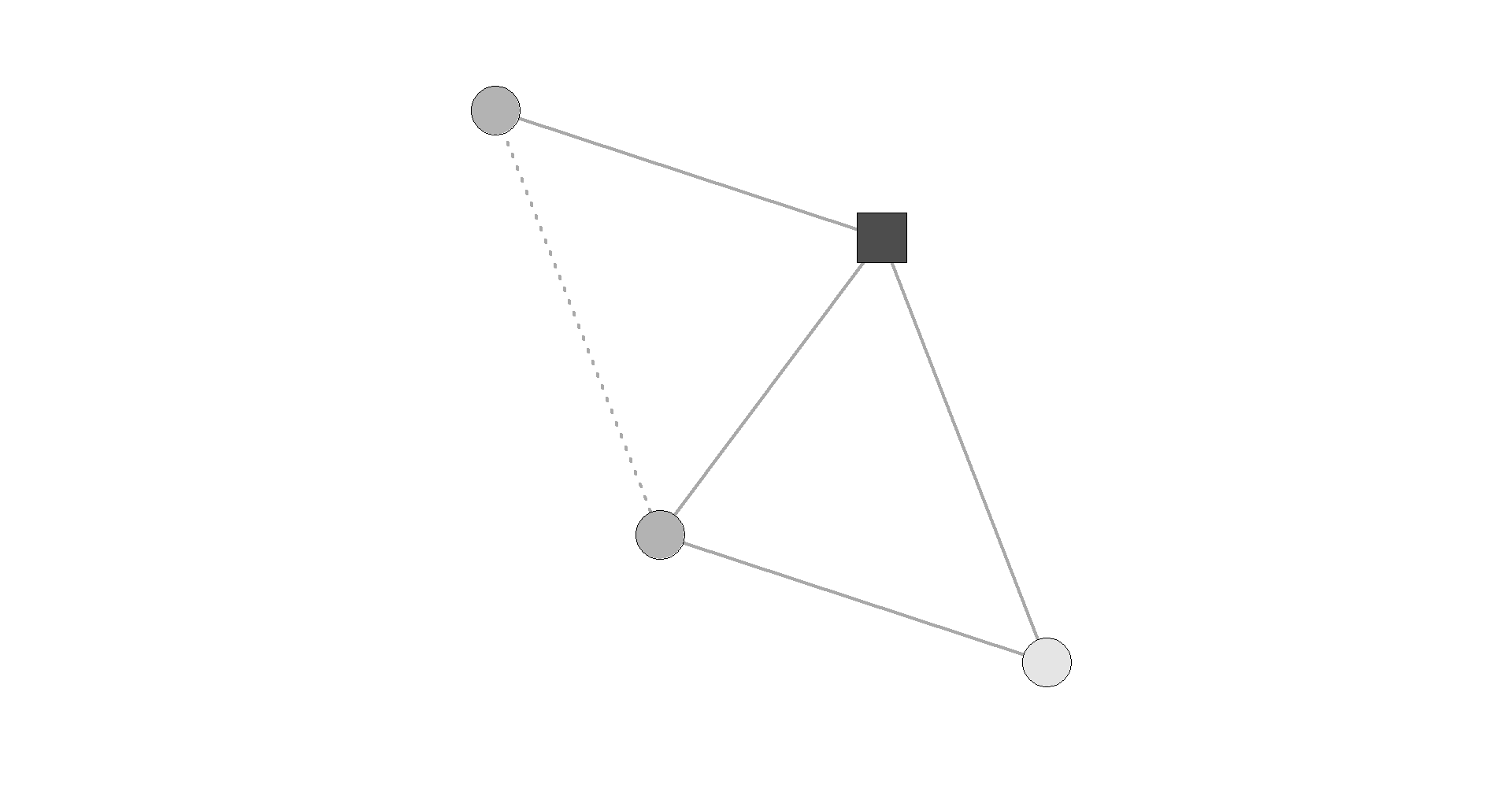}
  \caption{Sample operation 2}
  \label{fig:op2}
\end{subfigure}

\begin{subfigure}{.48\textwidth}
  \centering
  \includegraphics[width=\linewidth]{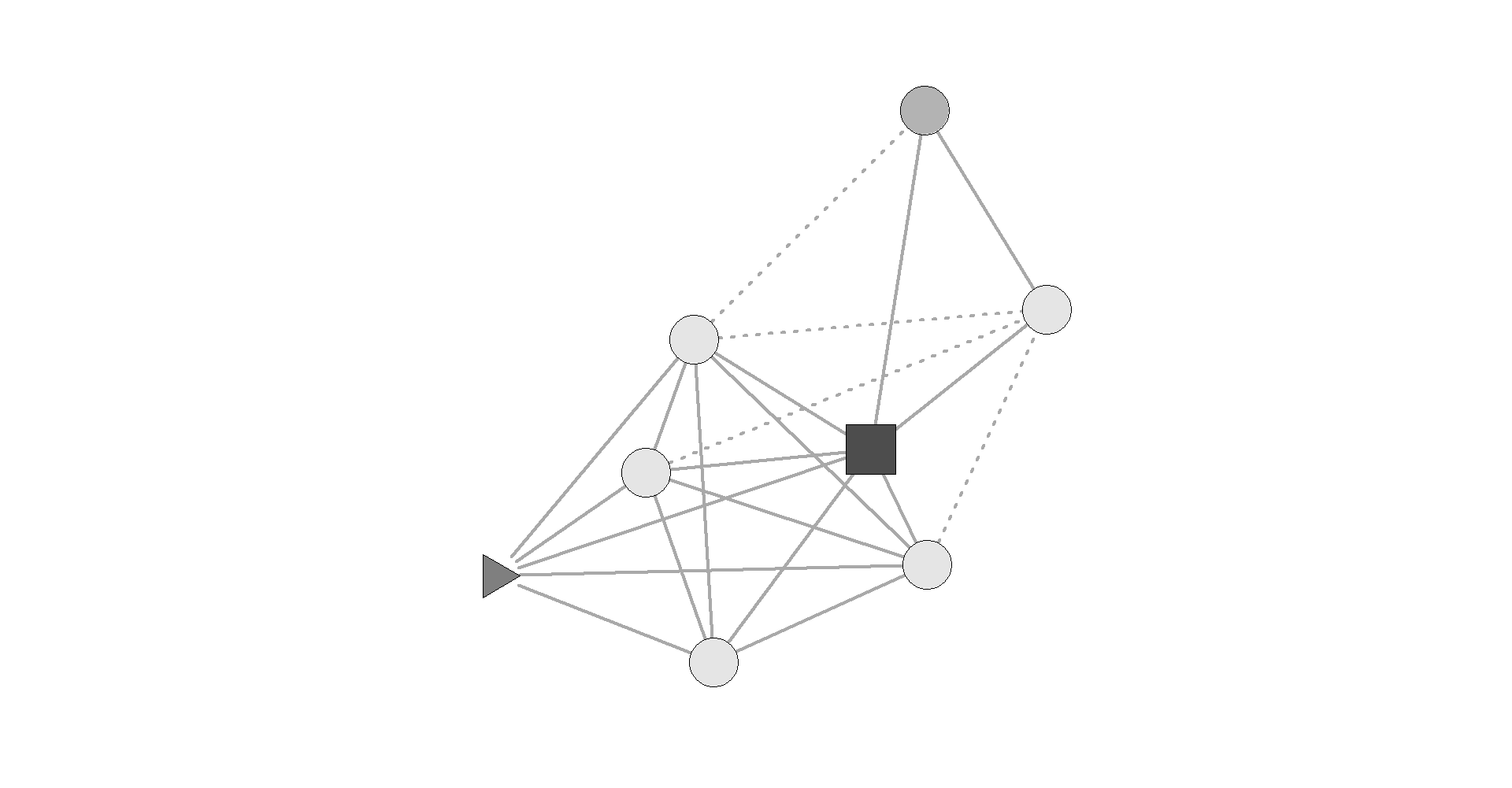}
  \caption{Sample operation 3}
  \label{fig:op3}
\end{subfigure}%
\begin{subfigure}{.48\textwidth}
  \centering
  \includegraphics[width=\linewidth]{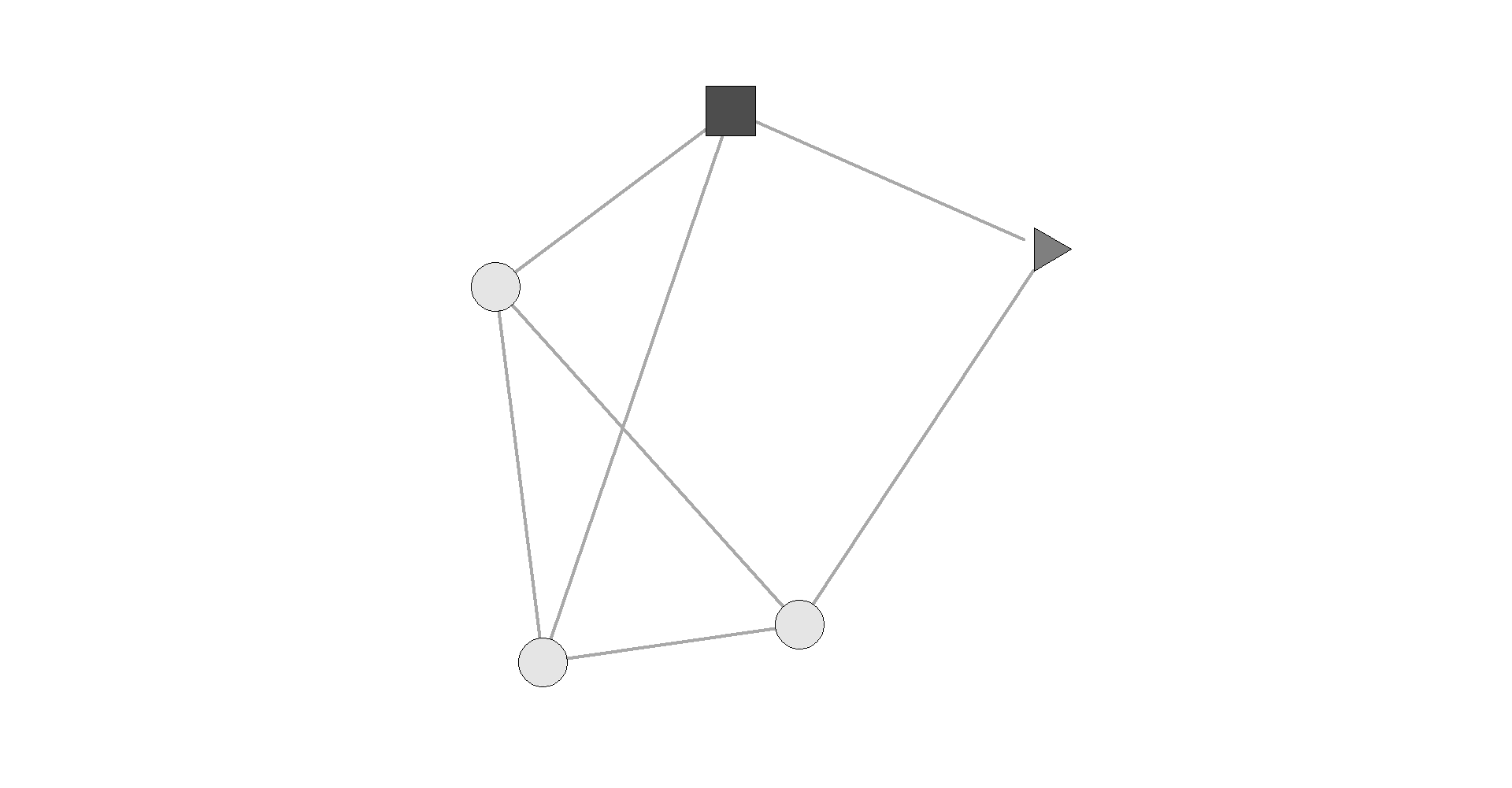}
  \caption{Sample operation 4}
  \label{fig:op4}
\end{subfigure}

\begin{subfigure}{.48\textwidth}
  \centering
  \includegraphics[width=\linewidth]{HT5_5_5alt.png}
  \caption{Sample operation 5}
  \label{fig:op5}
\end{subfigure}%
\caption{Sample operations within a generated human trafficking network}
\label{fig:sampleOps}
\end{figure}

\clearpage
\section{Results of Spectral Analysis}
\label{app:spec}
\begin{table}[h]
\centering
\caption{Comparison of spectra of Laplacian matrices of real and synthetic networks}
\begin{tabular}{|c|c|c|c|c|c|}
\hline
Operation & 1 & 2 & 3 & 4 & 5 \\ \hline\hline
1 & - & 0 & 8.428 & 2 & 7.348 \\ \hline
2 & - & - & 8.428 & 2 & 7.348 \\ \hline
3 & - & - & - & 7.557 & 2.248 \\ \hline
4 & - & - & - & - & 6.481 \\ \hline
5 & - & - & - & - & - \\ \hline\hline
AG1 & 1.414 & 1.414 & 9.381 & 3.162 & 8.246 \\ \hline
AG2.1 & 3.162 & 3.162 & 10.392 & 4.243 & 9.165 \\ \hline
AG2.2 & 0 & 0 & 8.428 & 2 & 7.348 \\ \hline
AG2.3 & 0 & 0 & 8.428 & 2 & 7.348 \\ \hline\hline
Central & 5.099 & 5.099 & 4.914 & 4.243 & 3.464 \\ \hline
Chalice & 3.972 & 3.972 & 4.977 & 2.938 & 4.279 \\ \hline
Retriever & 17.745 & 17.745 & 12.279 & 17.062 & 13.585 \\ \hline
Span & 3.309 & 3.309 & 5.638 & 2.115 & 4.559 \\ \hline
\end{tabular}

\label{tab:spec1}
\end{table}

\begin{table}[h]
\centering
\caption{Comparison of spectra of Laplacian matrices of real networks from different sources}
\begin{tabular}{|c|c|c|c|c|c|c|c|c|}
\hline
Operation & AG1 & AG2.1 & AG2.2 & AG2.3& Central & Chalice & Retriever & Span \\ \hline
AG1 & - & 2 & 1.414 & 1.414 & 6 & 5.251 & 18.579 & 4.559 \\ \hline
AG2.1 & - & - & 3.162 & 3.162 & 6.928 & 6.633 & 19.545 & 5.831 \\ \hline
AG2.2 & - & - & - & 0 & 5.099 & 3.973 & 17.745 & 3.309 \\ \hline
AG2.3 & - & - & - & - & 5.099 & 3.973 & 17.745 & 3.309 \\ \hline
Central & - & - & - & - & - & 3.008 & 15.295 & 2.753 \\ \hline
Chalice & - & - & - & - & - & - & 14.944 & 1.084 \\ \hline
Retriever & - & - & - & - & -& - & - & 15.642 \\ \hline
Span & - & - & - & - & -& - & - & - \\ \hline
\end{tabular}

\label{tab:spec2}
\end{table}

\clearpage
\section{Formal Model Description}
\label{app:model}
\spacingset{1.5}
We now define the necessary components for MFNIP-R. Let $x_{ij}$ be the amount of flow through arc $(i,j) \in A \cup A^{R,out}$, and $x_i$ be the amount of flow through node $i$. Let $y_i$ be the indicator of whether or not node $i$ has been interdicted, setting its capacity to $0$, with $Y$ being the set of feasible interdiction decisions. Let $z^{out}_{ij}$ be the indicator of restructuring arc $(i,j) \in A^{R,out}$ from node $i$, and let $z^{in}_{ij}$ be the indicator of restructuring arc $(i,j) \in A^{R,in}$ from $j$, with $Z^{out}(y)$ and $Z^{in}(y)$ being the sets of feasible restructuring decisions from $i$ or $j$, respectively, with respect to the interdiction decision $y$. We note that we restrict the set of restructuring decisions to be dependent on the implemented interdiction decisions because we are specifically interested in how the traffickers will react to the interdictions.

In this case study, instead of assigning a global budget for restructuring, we assign each trafficker a budget. We choose to limit the number of actions each trafficker can take, as the actions of one trafficker may not prevent the actions of the other traffickers. For ease of defining constraints, we further partition $A^{R,out}$ and $A^{R,in}$ into sets based on which trafficking operations are involved. Let $A^{R,out}_i$ be the set of restructurable ``out" arcs that can be restructured by trafficker $i$, and $A^{R, in}_i$ be the set of restructurable ``in" arcs that involve trafficker $i$. We note that these sets are disjoint between different traffickers, i.e. $A^{R,out}_i \cap A^{R,out}_j = \emptyset$ for $i,j \in T$ and $i \ne j$, and $A^{R,out} = \cup_{i\in T} A^{R,out}_i$.

Let $c_{jk}^i$ be the cost for trafficker $i$ to restructure the arc $(j,k)$, and let $b^i$ be the budget of trafficker $i$. For a fixed attacker solution $\bar{y}$, we can define the defender's problem as:

\begin{singlespace}
{\footnotesize\begin{align} % defender constraints
    \max_{x, z} ~~~ & \sum_{i \in N: (s,i) \in A \cup A^{R,out}} x_{si} & \nonumber\\
    \text{s.t. }& \sum_{(h,i) \in A \cup A^{R,out}} x_{hi} = x_i & \text{ for } i \in N\\
    & x_i = \sum_{(i,h) \in A \cup A^{R,out}} x_{ih} & \text{ for } i \in N \\
    & x_{ij} \le u_{ij} & \text{ for } (i,j) \in A \\
    & x_{ij} \le u_{ij} z^{out}_{ij} & \text{ for } (i,j) \in A^{R,out} \setminus A^{R,in} \\
    & x_{ij} \le u_{ij} (z^{out}_{ij}+z^{in}_{ij}) & \text{ for } (i,j) \in A^{R,in} \\
    & x_i \le u_i(1 - y_i) & \text{ for } i \in N\setminus\{j \in V: \exists h \in B \text{ with } (h,j) \in B^R \} \\
    & x_j \le u_j(1 - y_j) + \Tilde{u_j} z^{out}_{sj} & \text{ for } j \in V \text{s.t.} \exists i \in B, (i,j) \in B^R \\
    &\sum_{(j,k) \in A^{R,out}_i} c_{jk}^i z^{out}_{jk} + \sum_{(j,k) \in A^{R,in}_i} c_{jk}^i z^{in}_{jk} \le b^i & \text{ for } i \in T \\
    &\sum_{j \in V} z^{out}_{ij} \le \sum_{h \in V: (i,h) \in A} \bar{y}_h & \text{ for } i \in T\\
    &\sum_{i \in T} z^{in}_{ij} \le \sum_{h \in T \cup B: (h,j) \in A} \bar{y}_h & \text{ for } j \in V\\
    &\sum_{j \in V: (i,j) \in T^R} z^{out}_{sj} \le \bar{y}_i & \text{ for } i \in T \\
    &\sum_{j \in V: (i,j) \in B^R} z^{out}_{sj} \le \bar{y}_i & \text{ for } i \in B \\
    &z^{out}_{sj} \le 1 - \bar{y}_j & \text{ for } j \in V \text{ s.t. } \exists i \in B, (i,j) \in B^R \\
    &\sum_{k \in V: (j,k) \in B^R, (i,j) \in A} z^{out}_{sk} \le 1 & \text{ for } i \in T \\
    &\sum_{k \in V: (j,k) \in A^{R,out}_i} z^{out}_{jk} \le |\{k \in V: (j,k) \in A^{R,out}_i\}| z^{out}_{sj} & \text{ for } i \in T, j \in V \text{ s.t. } \exists h \in B, (h,j) \in B^R \\
    &\sum_{i \in T: (i,j) \in A^{R, out}_i \cap A^{R, in}_i} (z^{in}_{ij} + z^{out}_{ij}) \le 1 & \text{ for } j \in V\\
    &(z^{out}_{ij} + z^{in}_{ij}) \le 1 & \text{ for } (i,j) \in A^{R,out}_i \cap A^{R,in}_i\\
    & x \ge 0 & \\
    & z^{out} \in \{0,1\}^{|A^{R,out}|} & \\
    & z^{out} \in \{0,1\}^{|A^{R,in}|} & 
\end{align}}%
\end{singlespace}

The objective function of the MILP is maximizing the amount of flow out of the source node. Constraints (1)-(2) are flow balance constraints. Constraints (3)-(7) enforce the flow through an arc or node is at most the capacity of that arc or node, as adjusted by interdictions and restructurings. In particular, constraints (7) incorporate the decrease in flow through a victim node from interdiction and increase from being promoted to the role of bottom. We note that, for $(i,j) \in A^{R,in}$, $x_{ij}$ is bounded above by both $z^{out}_{ij}$ and $z^{in}_{ij}$. Additionally, since $A^{R,in} \subset A^{R,out}$, summing over $A \cup A^{R,out}$ in constraints (1)-(2) will include all relevant arcs. Constraints (8) enforce the restructuring budget for each trafficker. Constraints (9)-(10) allow for restructurings from the traffickers or bottoms based on the implemented interdictions. Note that the sum on the right-hand side of constraints (10) is for ease of determining which trafficking operation each victim is a part of. Constraints (11) allow for a back-up trafficker to be restructured to if the trafficker has been interdicted, and Constraints (12) allow for a victim to be promoted to bottom if the operation's bottom has been interdicted. Constraints (13) prevent an interdicted victim from being promoted to bottom, and constraints (14) enforce that at most one victim can be promoted to bottom. Constraints (15) allow for the trafficker to give victims to the newly promoted bottom. Constraints (16) prevent a victim from being recruited into more than one operation. For arcs in $A^{R,out}_i \cap A^{R,in}_i$, constraints (17) prevent both $z^{out}$ and $z^{in}$ from being nonzero. Constraints (18) enforce that the flow variables are non-negative, and constraints (19) and (20) enforce that the restructuring variables are binary.

The attacker is assigned a budget $b^a$ to interdict the network, limiting their ability to disrupt the network. Each node $i$ is assigned a cost $r_i$ that must be expended to interdict that node. In drug trafficking, ``climbing the ladder" constraints are used to emulate how law enforcement would pursue prosecuting cases against participants in the drug trafficking network \citep{malaviya2012multi}. In disrupting sex trafficking networks, it is not necessary to interdict victims in order to interdict traffickers; however the cooperation of victims can assist in the successful prosecution of traffickers. To model this, the cost of interdicting a trafficker node is reduced by a fixed amount based on whether or not their bottom has been interdicted, as well as the number of their victims that have been interdicted. Let $d_k^i$ be the decrease in interdiction cost of trafficker $i$ by interdicting node $k$. Additionally, let $r_i^{min}$ be the minimum the cost to interdict trafficker $i$, and define the variable $\Tilde{r}_i$ be the adjusted cost of interdicting trafficker $i$. We can define the attacker's problem as:

\begin{singlespace}
{\footnotesize\begin{align} % attacker constraints
    \min_{y, \Tilde{r}} \max_{x, z} ~~~ & \sum_{i \in N: (s,i) \in A \cup A^{R,out}} x_{si} & \nonumber\\
    \text{s.t. }& \text{ Constraints (1) - (20)} \nonumber \\
    &\sum_{i \in T} \Tilde{r}_i y_i + \sum_{j \in B \cup T} r_i y_i \le b^a & \\
    &\Tilde{r}_i \ge r_i^{min} & \text{ for } i \in T\\
    &\Tilde{r}_i \ge r_i - \sum_{k \in B \cup V} d_k^i y_i & \text{ for } i \in T \\
    & y \in \{0,1\}^{|N|} &
\end{align}}%
\end{singlespace}

Constraint (21) is the budget of the attacker. Constraints (22) and (23) adjust the cost to interdict the traffickers. Constraints (24) enforce that the interdiction variables are binary. We note that, as qualitative research on the operations of sex trafficking networks is furthered, this model can be refined to provide more accurate insights regarding disrupting sex trafficking networks.

Note that $\Tilde{r}_i y_i$ is a bilinear term. This can be linearized by replacing it with a new variable $\Tilde{y}_i$ and including the McCormick inequalities in the set of constraints. Since $y_i$ is binary, the McCormick inequalities will result in $\Tilde{y}_i = \Tilde{r}_i y_i$.

\clearpage
\section{Interdiction Results}
\label{app:moreResults}
\begin{figure}[h]
        \begin{subfigure}{.48\textwidth}
  \centering
  \includegraphics[width=.85\linewidth]{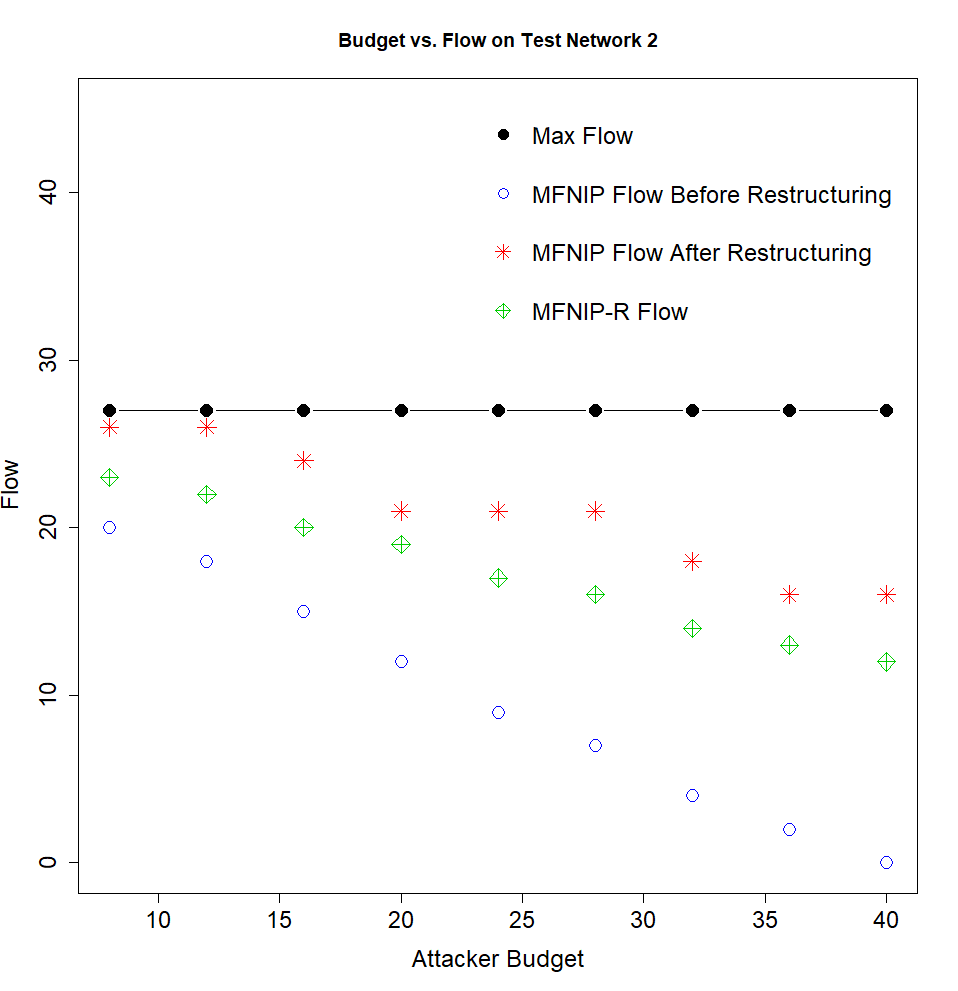}
  \label{fig:flow2}
\end{subfigure}%
\begin{subfigure}{.48\textwidth}
  \centering
  \includegraphics[width=.85\linewidth]{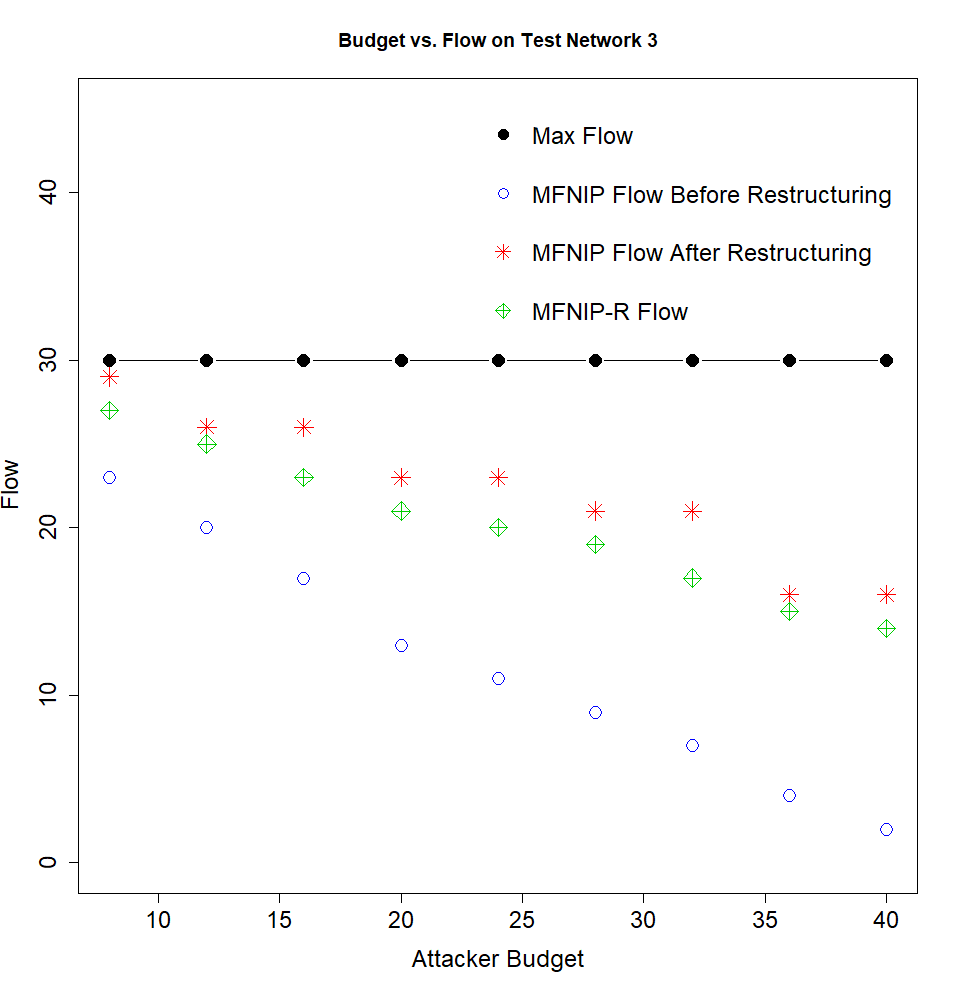}
  \label{fig:flow3}
\end{subfigure}

\begin{subfigure}{.48\textwidth}
  \centering
  \includegraphics[width=.85\linewidth]{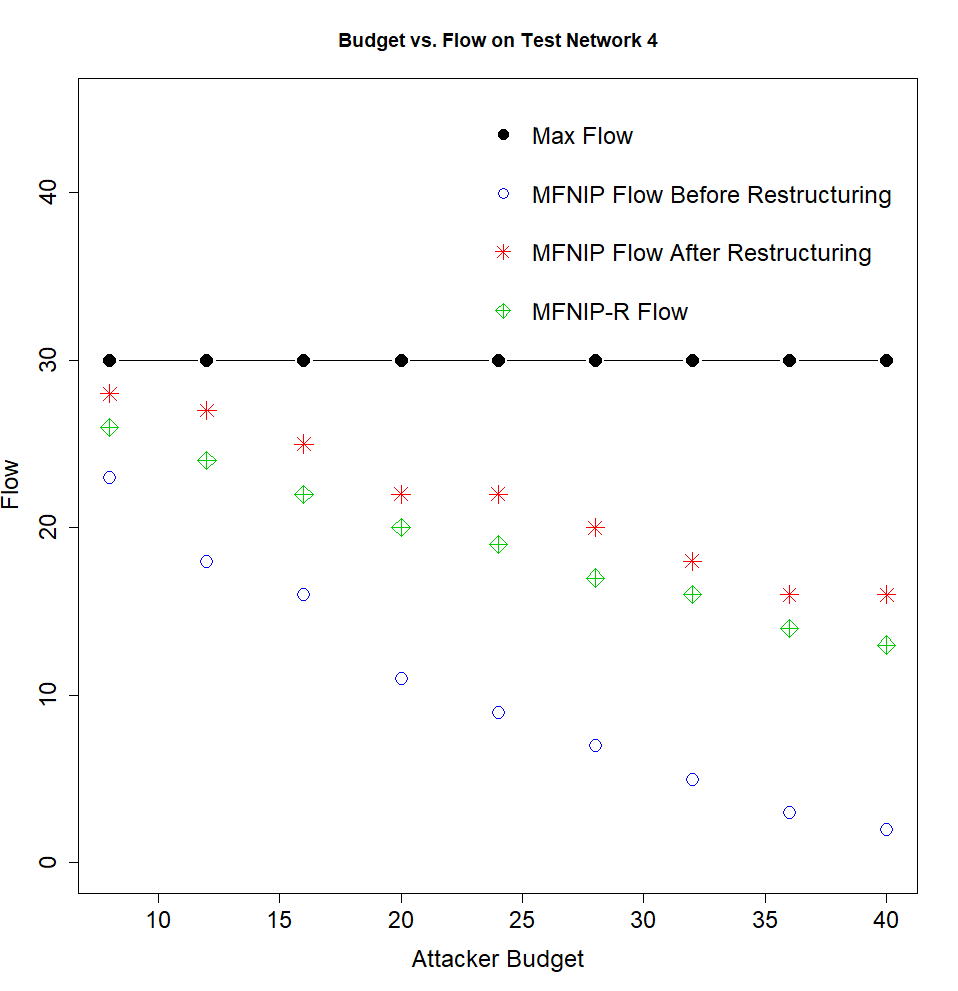}
  \label{fig:flow4}
\end{subfigure}%
\begin{subfigure}{.48\textwidth}
  \centering
  \includegraphics[width=.85\linewidth]{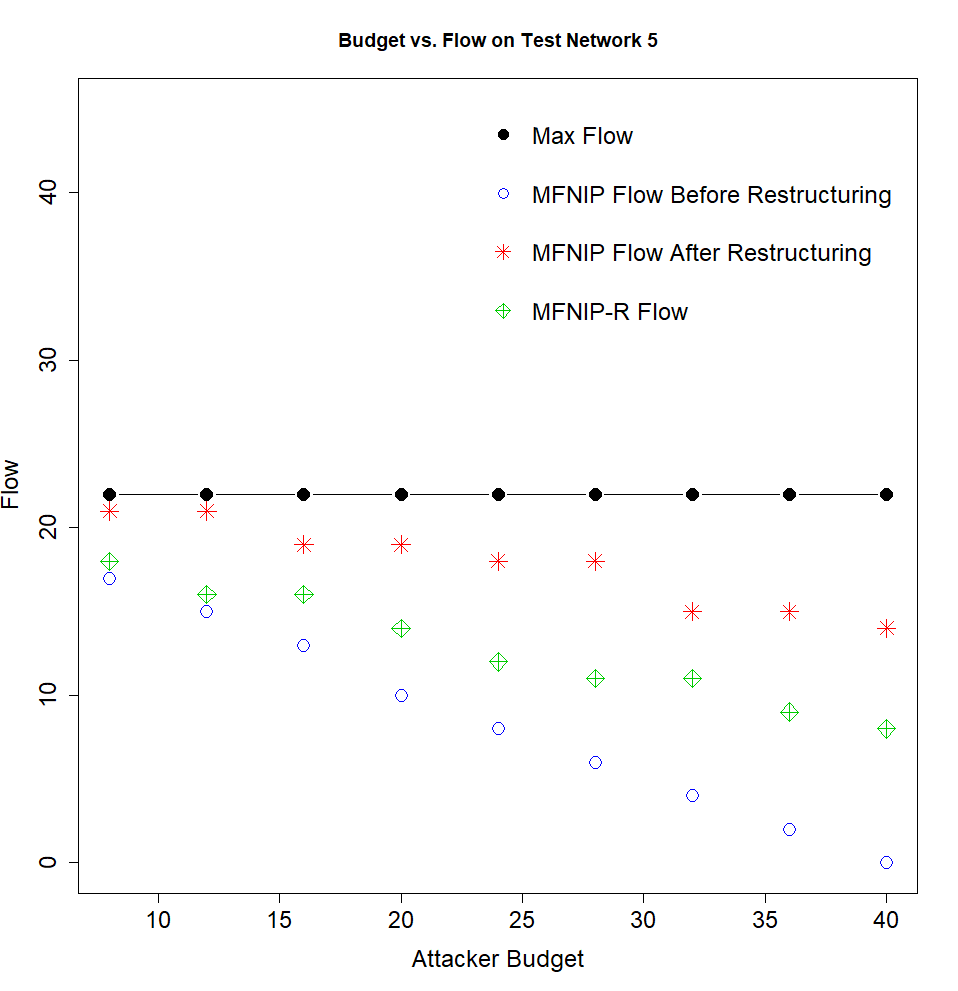}
  \label{fig:flow5}
\end{subfigure}%
\caption{MFNIP-R flows over varying attacker budgets}
\end{figure}

\begin{table}[h!]
    \centering
    \caption{Interdiction recommendations from MFNIP and MFNIP-R for Network 2}
    \begin{tabular}{|c|c|c|c|c|c|c|}
          \hline
         Attacker & MFNIP & MFNIP & MFNIP  & MFNIP-R & MFNIP-R & MFNIP-R \\
         Budget & Int Trafficker & Int Bottom & Int Victim & Int Trafficker & Int Bottom & Int Victim \\
         \hline
         8        & 0                                   & 2                               & 0                               & 0                                   & 0                               & 4                               \\ \hline
12       & 0                                   & 3                               & 0                               & 1                                   & 0                               & 4                               \\ \hline
16       & 0                                   & 2                               & 2                               & 1                                   & 0                               & 6                               \\ \hline
20       & 1                                   & 2                               & 2                               & 1                                   & 1                               & 6                               \\ \hline
24       & 2                                   & 3                               & 1                               & 1                                   & 1                               & 8                               \\ \hline
28       & 2                                   & 4                               & 1                               & 2                                   & 0                               & 10                              \\ \hline
32       & 3                                   & 4                               & 0                               & 2                                   & 0                               & 13                              \\ \hline
36       & 3                                   & 4                               & 2                               & 3                                   & 1                               & 9                               \\ \hline
40       & 3                                   & 4                               & 5                               & 3                                   & 2                               & 9                               \\ \hline

    \end{tabular}
    \label{tab:whoInt2}
\end{table}

\begin{table}[h!]
    \centering
    \caption{Interdiction recommendations from MFNIP and MFNIP-R for Network 3}
    \begin{tabular}{|c|c|c|c|c|c|c|}
          \hline
         Attacker & MFNIP & MFNIP & MFNIP  & MFNIP-R & MFNIP-R & MFNIP-R \\
         Budget & Int Trafficker & Int Bottom & Int Victim & Int Trafficker & Int Bottom & Int Victim \\
         \hline
        8        & 0                                   & 2                               & 0                               & 0                                   & 1                               & 2                               \\ \hline
12       & 1                                   & 1                               & 1                               & 1                                   & 0                               & 4                               \\ \hline
16       & 1                                   & 2                               & 1                               & 1                                   & 0                               & 6                               \\ \hline
20       & 2                                   & 2                               & 1                               & 2                                   & 1                               & 3                               \\ \hline
24       & 2                                   & 3                               & 1                               & 2                                   & 1                               & 6                               \\ \hline
28       & 2                                   & 2                               & 5                               & 3                                   & 3                               & 1                               \\ \hline
32       & 2                                   & 3                               & 5                               & 2                                   & 2                               & 8                               \\ \hline
36       & 4                                   & 4                               & 0                               & 3                                   & 3                               & 5                               \\ \hline
40       & 4                                   & 4                               & 2                               & 4                                   & 3     & 5 \\\hline
    \end{tabular}
    \label{tab:whoInt3}
\end{table}

\begin{table}[h!]
    \centering
    \caption{Interdiction recommendations from MFNIP and MFNIP-R for Network 4}
    \begin{tabular}{|c|c|c|c|c|c|c|}
          \hline
         Attacker & MFNIP & MFNIP & MFNIP  & MFNIP-R & MFNIP-R & MFNIP-R \\
         Budget & Int Trafficker & Int Bottom & Int Victim & Int Trafficker & Int Bottom & Int Victim \\
         \hline
        8        & 0                                   & 2                               & 0                               & 0                                   & 1                               & 2                               \\ \hline
12       & 1                                   & 1                               & 1                               & 1                                   & 0                               & 4                               \\ \hline
16       & 1                                   & 2                               & 1                               & 1                                   & 0                               & 6                               \\ \hline
20       & 2                                   & 2                               & 1                               & 1                                   & 0                               & 8                               \\ \hline
24       & 2                                   & 3                               & 1                               & 2                                   & 1                               & 5                               \\ \hline
28       & 2                                   & 2                               & 5                               & 2                                   & 1                               & 7                               \\ \hline
32       & 3                                   & 3                               & 2                               & 3                                   & 2                               & 5                               \\ \hline
36       & 3                                   & 4                               & 2                               & 3                                   & 3                               & 5                               \\ \hline
40       & 3                                   & 4                               & 5                               & 4                                   & 3                               & 5                               \\ \hline
    \end{tabular}
    \label{tab:whoInt4}
\end{table}

\begin{table}[h!]
    \centering
    \caption{Interdiction recommendations from MFNIP and MFNIP-R for Network 5}
    \begin{tabular}{|c|c|c|c|c|c|c|}
          \hline
         Attacker & MFNIP & MFNIP & MFNIP  & MFNIP-R & MFNIP-R & MFNIP-R \\
         Budget & Int Trafficker & Int Bottom & Int Victim & Int Trafficker & Int Bottom & Int Victim \\
         \hline
        8        & 0                                   & 2                               & 0                               & 0                                   & 0                               & 4                               \\ \hline
12       & 0                                   & 2                               & 2                               & 0                                   & 0                               & 6                               \\ \hline
16       & 1                                   & 2                               & 2                               & 0                                   & 0                               & 8                               \\ \hline
20       & 2                                   & 2                               & 1                               & 0                                   & 0                               & 10                              \\ \hline
24       & 2                                   & 2                               & 3                               & 1                                   & 0                               & 10                              \\ \hline
28       & 2                                   & 3                               & 3                               & 1                                   & 1                               & 10                              \\ \hline
32       & 2                                   & 3                               & 5                               & 1                                   & 0                               & 13                              \\ \hline
36       & 2                                   & 4                               & 5                               & 2                                   & 1                               & 12                              \\ \hline
40       & 2                                   & 4                               & 7                               & 2                                   & 1                               & 13                              \\ \hline
    \end{tabular}
    \label{tab:whoInt5}
\end{table}
\end{appendices}

\end{document}